\documentclass[aps,prb,showpacs,twoside,amsmath,amssymb,floatfix,superscriptaddress,,twocolumn,fleqn]{revtex4}
\usepackage{graphicx}
\usepackage{txfonts}
\usepackage{bm}
\DeclareMathOperator{\sgn}{sgn}
\DeclareMathOperator{\dotproduct}{\cdot}
\DeclareMathOperator{\Imag}{Im}

\usepackage[bookmarks=true,colorlinks=true,urlcolor=blue,linkcolor=blue,citecolor=blue,breaklinks]{hyperref} 
\usepackage[usenames,dvipsnames]{color}

\begin{document}
\newcommand{\brho}{{\bm \rho}}
\def \beps{{\hbox{\boldmath $\epsilon$}}}
\def \bdelta{{\hbox{\boldmath $\delta$}}}
\pacs{73.22.Pr, 73.22.Lp}

\title{Quasiparticle spectra and excitons of organic molecules deposited on substrates:\\$\textit{G}_{\text{0}}\textit{W}_{\text{0}}$-BSE approach applied to benzene on graphene and metallic substrates}
\author{V. Despoja}
\email{vito@phy.hr}
\affiliation{Department of Physics, University of Zagreb, Bijeni\v{c}ka 32, HR-10000 Zagreb, Croatia}
\affiliation{Donostia International Physics Center (DIPC), Paseo de Manuel de Lardizabal 4, ES-20018 San Sebast{\'{i}}an, Spain}
\author{I. Lon\v cari\' c}
\affiliation{Department of Physics, University of Zagreb, Bijeni\v{c}ka 32, HR-10000 Zagreb, Croatia}
\author{D. J. Mowbray}
\email{duncan.mowbray@gmail.com}
\affiliation{Donostia International Physics Center (DIPC), Paseo de Manuel de Lardizabal 4, ES-20018 San Sebast{\'{i}}an, Spain}
\affiliation{Nano-Bio Spectroscopy Group and ETSF Scientific Development Center, Departamento de F{\'{\i}}sica de Materiales, Universidad del 
Pa{\'{\i}}s Vasco UPV/EHU, ES-20018 San Sebasti{\'{a}}n, Spain}
\author{L. Maru\v si\' c}
\affiliation{Maritime Department, University of Zadar, M. Pavlinovi\'{c}a b.b., HR-23000 Zadar, Croatia}
\begin{abstract}
We present an alternative methodology for calculating the quasi-particle energy, energy loss, and 
optical spectra of a molecule deposited on graphene or a metallic substrate. To test the accuracy of the method it is first 
applied to the isolated benzene ($\text{C}_6\text{H}_6$) molecule. The quasiparticle energy levels and especially the energies of the benzene excitons (triplet, singlet, optically active and inactive) are in very good agreement with available experimental 
results. It is shown that the vicinity of the various substrates (pristine/doped graphene or (jellium) metal surface) reduces the quasiparticle HOMO--LUMO gap by an amount that slightly depends on the substrate type. This is consistent with the simple 
image theory predictions. It is even shown that the substrate does not change the energy of the excitons in the isolated molecule. We prove (in terms of simple image theory) that energies of the excitons are indeed influenced by two 
mechanisms which cancel each other. We demonstrate that the benzene singlet optically active ($E_{1u}$) exciton 
couples to real electronic excitations in the substrate. This causes it substantial decay, such as $\Gamma \approx 174$~meV for 
pristine graphene and $\Gamma \approx 362$~meV for metal surfaces as the substrate. However, we find that doping graphene does not influence the $E_{1u}$ exciton decay rate. 
\end{abstract}

\maketitle

\section{Introduction}
Nowadays, $\pi$-conjugated organic molecules and their derivative films \cite{pentacenethinfilms} are increasingly used in many applications. They are often used in organic electronic devices such as field effect transistors \cite{orgFET} or organic transistors.\cite{orgtranz} Also, their good charge mobility and small optical band gap makes these materials suitable for photovoltaic applications, such as in solar cells.\cite{orgfotovol} Moreover, the spatial localization of organic molecules allows the light absorbed by the molecule to be converted into substrate excitations such as surface plasmon or electron hole excitations. The latter could be of interest in biosensing applications.\cite{biosensors} 
This has spurred recent studies characterizing the formation via cyclization cascade reactions \cite{DuncanScience} and the charge transfer within combined donor--acceptor layers \cite{DuncanACSNano} of $\pi$-conjugated organic molecules on metal substrates.
However, the basic building block of the most utilized organic molecules, such as aromatic hydrogen carbonates, is the benzene ring.   

This work is motivated by all these potential applications, and is focused on exploring the quasiparticle and optical properties of benzene deposited on semimetallic (pristine graphene) and various metallic (doped graphene and Ag(jellium) surface) substrates. Special attention is paid to examine the influence of the substrate on the molecular HOMO--LUMO gap, exciton plasmon interaction, efficiency of the molecule mediated light substrate plasmon conversion, and the decay of the molecular excitons to electron-hole excitations in the substrate.

 In the formulation of the problem we shall use previous theories which are well established and have been tested by various spectroscopic experiments.\cite{StevenLouie2006,Spectro3,BSE-Benzene} The quasiparticle properties of the deposited molecule will be investigated in the framework of Hedins's $GW$ theory \cite{StevenLouie2006,Hedin,GWtheory} while the optical properties will be investigated by solving the Bethe-Salpeter equation (BSE) whose practical application was first developed by Strinati,\cite{Strinati} and more recently by Louie \emph{et al.}\cite{StevenLouie2006,StevenLouie98,StevenLouie2000,Rubio}  These methodologies have been successfully used to calculate the electronic HOMO--LUMO gaps \cite{JuanmaRenormalization1} and optical gaps \cite{JuanmaRenormalization2} of benzene on various substrates.  However, the computational complexity of such calculations has limited their range of applicability, resulting in a need for simple models and benchmarking of the various levels of approximation.  For example, recent studies have shown that quasiparticle corrections to energy levels are linearly correlated with the fractions of the levels' densities within the substrate, molecule, and vacuum,\cite{Annapaola} with the quasiparticle gap renormalization proportional to the molecule's height above the substrate.\cite{Catalin}

However, to address these issues, the theoretical method presented here has been modified. Specifically, the optical spectra is obtained directly from the imaginary part of the dynamical 4-point polarizability matrix $L^{kl}_{ij}(\omega)$, which is the solution of the matrix BSE. In the standard resonant approximation \cite{StevenLouie98,StevenLouie2000,Rubio} the BSE reduces to an eigenvalue problem and the optical spectra is obtained in terms of BSE eigenvalues and eigenvectors. Moreover, we develop the methodology to examine dark excitons, i.e.\ excitons that cannot be excited by an external electromagnetic field. They can be seen as the energy loss of an external dipole driving the molecule, and its intensity also can be expressed in term of the imaginary part of the polarizability matrix $L^{kl}_{ij}(\omega)$. We also show that the equilibrium molecule/substrate separation is large enough so that their electronic densities do not overlap. As a consequence of this, the only modification which has to be done in the formulation, after the substrate is introduced, is to extend the bare Coulomb interaction propagator $V$ by the substrate induced Coulomb interaction $V+\Delta W$. Because of all these different approaches we have rewritten all previous expressions to make it clear where our approaches separate.
 
The theory is first developed generally and then applied to an isolated molecule and to a molecule deposited on various different substrates. To check the accuracy of our method, we compare the results obtained for isolated benzene with available experimental results. The calculated ionization energies and HOMO--LUMO gap for isolated benzene are in very good agreement with experimental results.\cite{Spectro4, Spectro2, exp_affin} Also, the calculated energies of benzene excitons (dark, bright, triplet and singlet) agree remarkably well (within $100$~meV) with experimental data.\cite{triplet1_3_vacuum, triplet4_vacuum, singlet_vacuum} We show that the introduction of the substrates reduces the HOMO--LUMO gap for about $2$~eV, and that this reduction weakly depends on the type of the substrate. A somewhat surprising result is that energies of the benzene excitons are barely affected by 
the presence of the substrate. This is because the substrate reduces the HOMO--LUMO gap which reduces the exciton energy, but at the same time the substrate weakens the excited electron-hole interaction which increases the exciton energy. We find that these two effect almost exactly cancel and exciton energies remains practically unchanged. We also find that all these effects can be simply explained by applying image theory to the screening of the electron and the hole, as was also theoretically observed in Ref.~\onlinecite{StevenLouie2006}.

 The spectra of molecular excitations is obtained in such a way that an external probe (electromagnetic wave or dipole) can excite only excitations in the molecule but not in the substrate. This enables us to analyze the molecular spectra as spectra of driven/damped harmonic oscillator where the external probe is the driving force of frequency $\omega$, exciton in the molecule is a harmonic oscillator of frequency $\omega_0$, and the substrate is the source of damping with damping constant $\Gamma$. We find that only the singlet $E^1_{1u}$ exciton (bright exciton) decays when the molecule is in the vicinity of the substrate, while all other excitons fail to couple with the substrate and remain infinitely sharp. In the vicinity of graphene the $E^1_{1u}$ exciton decays into a continuum of $\pi-\pi^*$ interband electron-hole excitations where $\Gamma \approx 176$~meV. In the vicinity of the Ag(jellium) surface the $E^1_{1u}$ exciton decays faster and $\Gamma \approx 362$~meV. We also find that there are no extra peaks in any molecular spectra.  This means that the excitons do not interact with the 2D plasmon in doped graphene, or with the surface plasmon on the Ag(jellium) surface.

In Sec.~\ref{generaltheo} we present the general methodology used to solve the BSE for the 4-point polarizability matrix $L^{kl}_{ij}(\omega)$ and explain how to use the imaginary part of $L^{kl}_{ij}(\omega)$ to obtain the optical absorption and energy loss spectra for an arbitrary system.

In Sec.~\ref{Apptomol} the developed formulation is applied to derive the quasiparticle spectra, optical absorption and energy loss spectra of an isolated benzene molecule. In order to calculate quasiparticle spectra, optical absorption and energy loss spectra of the molecule near a substrate, we demonstrate that we only need to replace the bare Coulomb interaction $V$ with $V+\Delta W$, where $\Delta W$ is the substrate induced Coulomb interaction. 

In Sec.~\ref{Results} we present results where the developed theoretical formulation is first used to calculate the ionization energies and HOMO--LUMO gap in the isolated benzene molecule, and then the formulation is used to obtain the exciton energies and spectra of excitations in benzene deposited on various substrates. This is followed by concluding remarks in Sec.~\ref{Conclusions}.

\section{Formulation of the problem}
\label{generaltheo}

In this section we shall first present the general method we use to solve the Bethe-Salpeter (BS) equation for the 4-point polarizability $L({\bf r}_1,{\bf r}_2;{\bf r}'_1,{\bf r}'_2,\omega)$ which is also called the two-particle correlation function.\cite{Strinati} We shall then present how to obtain the optical absorption and energy loss spectra for an arbitrary system from $L$.

\subsection{General theory}
 
In absorption experiments a photon creates an electron and a hole. In the lowest approximation we can consider them as two independent particles, which leads to infinitely long lived electron-hole pairs that can be described as a product of two one-particle Green's functions. However, in reality, the situation is much more complex. Because of the electron-electron interaction the excited electron and hole can 
interact with other excitations in the molecule or they can annihilate or interact mutually. These are all responsible for the creation of their bound states called excitons. Therefore, to give a realistic description of optical absorption phenomena we have to calculate the two-particle Green's function $G_2$. However, since in the Dyson's expansion of $G_2$ there are two possible annihilations leading to independent electron-hole motion, one of them should be subtracted from $G_2$.  In this way, the quantity describing the propagation of the coupled electron-hole pair is defined as \cite{Strinati} 
\begin{equation}
L(1,2;1',2')=iG_2(1,2;1',2')-iG(1,1')G(2,2'),
\label{fourpoint}
\end{equation}
where 
\begin{equation}
G_2(1,2;1',2')=(-i)^2\left\langle T\left\{\Psi(1)\Psi(2)\Psi^{\dagger}(2')\Psi^{\dagger}(1')\right\}\right\rangle,
\end{equation}
is the exact two particle Green's function and  
\begin{equation}
G(1,2)=-i\left\langle T\left\{\Psi(1)\Psi^{\dagger}(2)\right\}\right\rangle,
\end{equation}
is the exact one particle Green's function. Each argument in (\ref{fourpoint}) represents a four-vector, e.g.\ $1\equiv({\bf r}_1,t_1)$.

\begin{figure}
\centering
\includegraphics[width=\columnwidth]{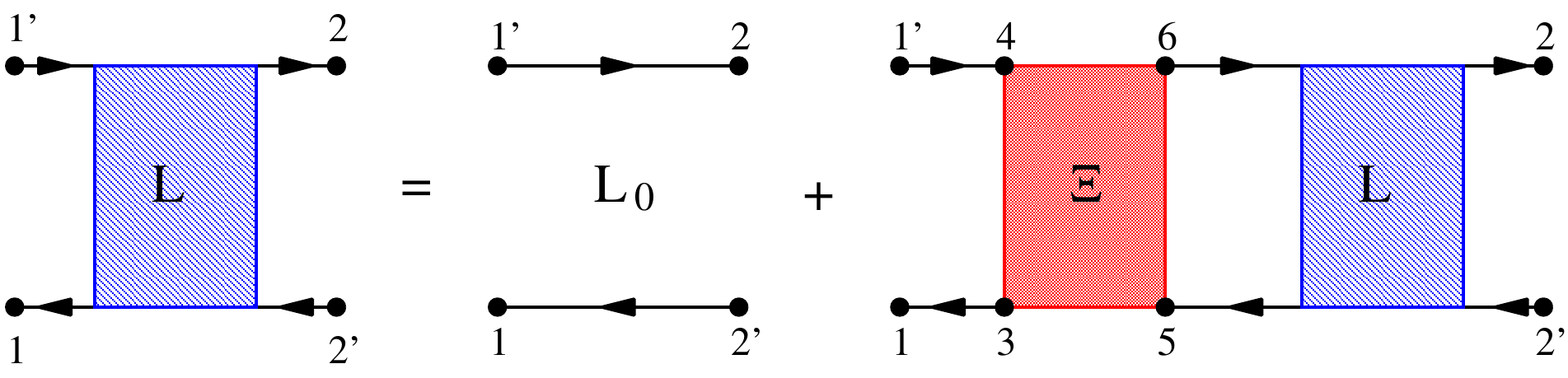}
\caption{(Color online) Bethe-Salpeter equation for general kernel $\Xi$.}
\label{Fig1}
\end{figure}

The 4-point polarizability (\ref{fourpoint}) satisfies a Dyson-like equation of the form \cite{Hedin,Strinati,GWtheory,Rubio} 
\begin{eqnarray}
L(1,2;1',2')&\!\!\!=\!\!\!&L_0(1,2;1',2')+
\nonumber\\
\nonumber
&\!\!\!&\!\int\! d3456 L_0(1,4;1',3)\Xi(3,6;4,5)L(5,2;6,2'),\\
\label{BSE}
\end{eqnarray}
also known as the Bethe-Salpeter equation, and shown in Feynman diagrams in Fig.~\ref{Fig1}.
The noninteracting 4-point polarizability has the form  
\begin{equation}
L_0(1,2;1',2')=-iG(1,2')G(2,1'),
\label{fourpoint0}
\end{equation}
and
\begin{equation}
\Xi(3,6;4,5)= i \frac{\delta}{\delta G(5,6)}\left\{V_H(3)\delta(3,4)+\Sigma_{\textit{XC}}(3,4)\right\},
\label{BSEkar}
\end{equation}
represents the Bethe-Salpeter kernel. Here 
\begin{equation}
V_H(3)=-i\int d1 V(3-1)G(1,1^+) 
\end{equation}
represents the exact Hartree energy and $\Sigma_{\textit{XC}}(3,4)$ represents the exact exchange-correlation self-energy, where the four-vector $1^+\equiv ({\bf r}_1, t_1+\delta)$ as $\delta \rightarrow 0^+$.

To calculate the exact kernel (\ref{BSEkar}) we have to know the exact self-energy $\Sigma_{\textit{XC}}$. However, this is not possible, so we have to make some approximations. The most frequently used approximation (in order to determine the BSE kernel) is the static-$GW$ approximation \cite{Rubio}
\begin{equation}
\Sigma_{\textit{XC}}(3,4)=iG(3,4)W({\bf r}_4,{\bf r}_3,\omega=0)\delta(t_4-t^+_3),
\end{equation}
where $W$ represents the exact statically screened Coulomb interaction. If we assume that $W$ weakly depends on $G$ then  
\begin{equation}
\frac{\delta W}{\delta G}\approx 0.
\end{equation}
The functional derivative in (\ref{BSEkar}) can be performed analytically, and the BSE kernel becomes
\begin{eqnarray}
\Xi(3,6;4,5)&=&V(3-5)\delta(3,4)\delta(5,6)
\nonumber\\
&&-W({\bf r}_4,{\bf r}_3,\omega=0)\delta(3,5)\delta(4,6)\delta(t_4-t_3).
\label{BSEkarTDSHFA}
\end{eqnarray}

Here we note that this static approximation is only justified when the frequency of the characteristic collective modes in the system are high enough to instantly screen the charge density fluctuation caused by electron-electron or hole-hole scattering in the system. Even though this is not always justified, we will use the approximation (\ref{BSEkarTDSHFA}) because it enables us to transform equation (\ref{BSE}) into frequency ($\omega$) 
space. 

More specifically, if we assume that the electron and hole are created and annihilated simultaneously, we can put $t_1'=t_1$, $t_2'=t_2$, and after using the approximation (\ref{BSEkarTDSHFA}), the 4-point polarizabilities $L$ and $L_0$ always appear as functions of two times $L(t_1,t_2)$ and $L_0(t_1,t_2)$. Moreover, due to the translational invariance in time, they become functions of the time difference $L(t_1-t_2)$ and $L_0(t_1-t_2)$. Using these properties and the fact that the BSE kernel is time independent, the Bethe-Salpeter equation can be Fourier transformed into $\omega$ space and becomes: 
\begin{widetext}
\begin{equation}
L({\bf r}_1,{\bf r}_2;{\bf r}'_1,{\bf r}'_2,\omega)=
L_0({\bf r}_1,{\bf r}_2;{\bf r}'_1,{\bf r}'_2,\omega)+\int d{\bf r}_3{\bf r}_4{\bf r}_5{\bf r}_6
L_0({\bf r}_1,{\bf r}_4;{\bf r}'_1,{\bf r}_3,\omega)
\Xi({\bf r}_3,{\bf r}_6;{\bf r}_4,{\bf r}_5)
L({\bf r}_5,{\bf r}_2;{\bf r}_6,{\bf r}'_2,\omega),
\label{BSEomega}
\end{equation} 
\end{widetext}
where the BSE kernel has the form \cite{Rubio}
\begin{eqnarray}
\Xi({\bf r}_3,{\bf r}_6;{\bf r}_4,{\bf r}_5)&=&
V({\bf r}_3,{\bf r}_5)\delta({\bf r}_3,{\bf r}_4)\delta({\bf r}_5,{\bf r}_6)\nonumber\\&&-
W({\bf r}_4,{\bf r}_3,\omega=0)\delta({\bf r}_3,{\bf r}_5)\delta({\bf r}_4,{\bf r}_6).
\label{BSEkerom}
\end{eqnarray}

\begin{figure}
\centering
\includegraphics[width=\columnwidth]{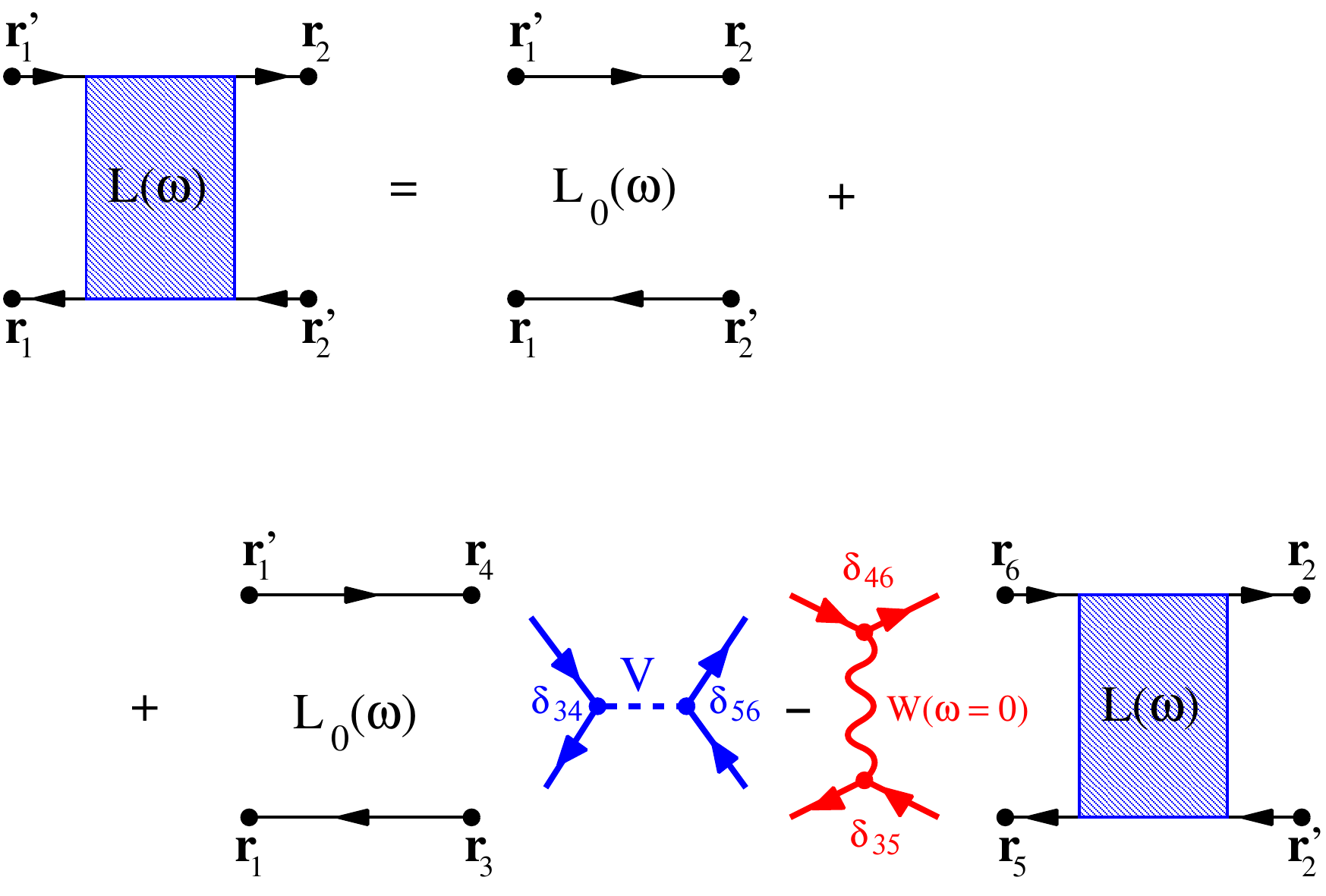}
\caption{(Color online) Bethe-Salpeter equation in Time Dependent Screened Hartree Fock (TDSHF) approximation.\cite{Rubio}}
\label{Fig2}
\end{figure}
The first term in (\ref{BSEkerom}) is usually called the BSE-Hartree kernel, and the second term is called  the BSE-Fock kernel. The Bethe-Salpeter equation (\ref{BSEomega}) is shown in Feynman diagrams in Fig.~\ref{Fig2}. The noninteracting 4-point polarizability $L_0$ then becomes

\begin{equation}
L_0({\bf r}_1,{\bf r}_2;{\bf r}'_1,{\bf r}'_2,\omega)=
-i\int^{\infty}_{-\infty}\frac{d\omega'}{2\pi} G({\bf r}_2,{\bf r}'_1,\omega')
G({\bf r}_1,{\bf r}'_2,\omega+\omega')
\label{freefor}
\end{equation}
and is shown by the first Feynman diagram on the right hand side in Fig.~\ref{Fig2}. 

\begin{figure}[h]
\centering
\includegraphics[width=0.4\columnwidth]{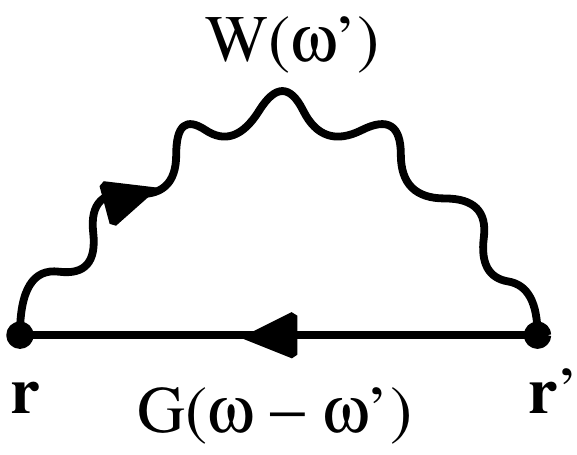}
\caption{$GW$ approximation for $\Sigma_{\textit{XC}}$.}
\label{Fig3}
\end{figure}

The Green's functions in (\ref{freefor}) can be obtained by solving the Dyson equation 
\begin{eqnarray}
G({\bf r},{\bf r}',\omega)&=&G_0({\bf r},{\bf r}',\omega)
\nonumber\\
\nonumber
&&+\int d{\bf r}_1{\bf r}_2 G_0({\bf r},{\bf r}_1,\omega)\Sigma({\bf r}_1,{\bf r}_2,\omega)G({\bf r}_2,{\bf r}',\omega),\\
\label{dysonG}
\end{eqnarray}
where $G_0$ is the independent electron Green's function and the self-energy can be separated into the Hartree part plus the exchange-correlation part
\begin{equation}
\Sigma=V_H+\Sigma_{\textit{XC}}.
\end{equation}
The Hartree term is  
\begin{equation}
V_H({\bf r},{\bf r}')=-i\int d{\bf r}_1G({\bf r}_1,{\bf r}_1,t,t^+)V({\bf r}_1,{\bf r})\delta({\bf r}-{\bf r}'),
\label{Hartree}
\end{equation}
and the exchange-correlation self energy term in the $GW$ approximation \cite{Hedin,GWtheory} reduces to 
\begin{equation}
\Sigma_{\textit{XC}}({\bf r},{\bf r}',\omega)=i\int^{\infty}_{-\infty}\frac{d\omega'}{2\pi}e^{-i\omega'\delta} 
G({\bf r},{\bf r}',\omega-\omega')W({\bf r},{\bf r}',\omega'),
\label{GWapprox}
\end{equation}
as shown in Feynman diagrams in Fig.~\ref{Fig3}.  

\begin{figure}[h]
\centering
\includegraphics[width=\columnwidth]{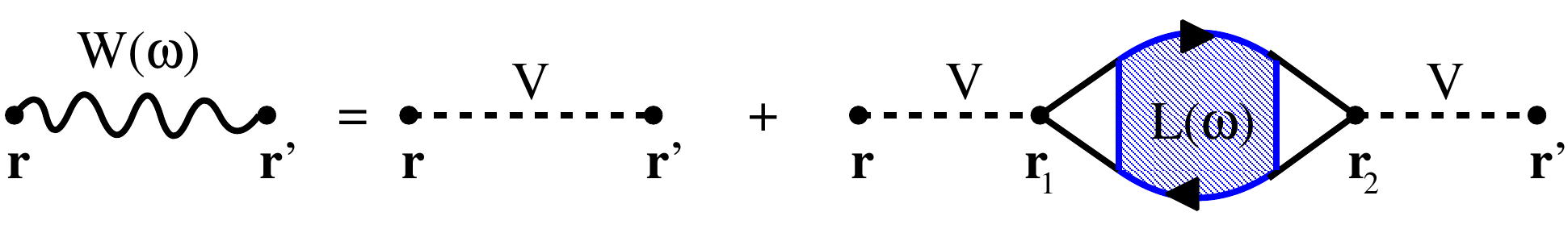}
\caption{(Color online) Propagator of the dynamically screened Coulomb propagator obtained from the polarizability $L$.}
\label{Fig4}
\end{figure}

The first quantity that needs to be obtained to solve this complicated set of equations is the propagator $W$ of the screened Coulomb interaction, since it appears in 
the BSE kernel (\ref{BSEkerom}) and is essential for the $GW$ approximation (\ref{GWapprox}). It is the solution of the equation      
\begin{eqnarray}
W({\bf r},{\bf r}',\omega)&=&V({\bf r},{\bf r}',\omega)
\nonumber\\
&&+\int d{\bf r}_1{\bf r}_2 V({\bf r},{\bf r}_1,\omega)\chi({\bf r}_1,{\bf r}_2,\omega)V({\bf r}_2,{\bf r}'),
\label{screened}
\end{eqnarray}
where the response function $\chi$ can be obtained from the 4-point polarizability by the coordinate annihilation    
\begin{equation}
\chi({\bf r}_1,{\bf r}_2,\omega)=L({\bf r}_1,{\bf r}_2;{\bf r}_1,{\bf r}_2,\omega). 
\label{chi}
\end{equation}
An equation for $W$ is shown in Feynman diagrams in Fig.~\ref{Fig4}. 

Therefore, equations (\ref{BSEomega}--\ref{chi}) form a self-consistent scheme, and we shall next describe the method we use to solve it. 

The first step is to solve the Density Functional Theory Kohn-Sham (DFT-KS) equations for the system in order to obtain the Kohn-Sham (KS) orbitals $\psi_i({\bf r})$ and energy levels $\varepsilon_i$. 
Using these KS states we can construct the independent electron Green's function 
\begin{equation}
G_0({\bf r},{\bf r}',\omega)=\sum_i\frac{\psi_i({\bf r})\psi^*_i({\bf r}')}{\omega-\varepsilon_i+i\eta\sgn(\varepsilon_F-\varepsilon_i)}.
\label{KSgreen}
\end{equation}

\begin{figure}[h]
\centering
\includegraphics[width=0.4\columnwidth]{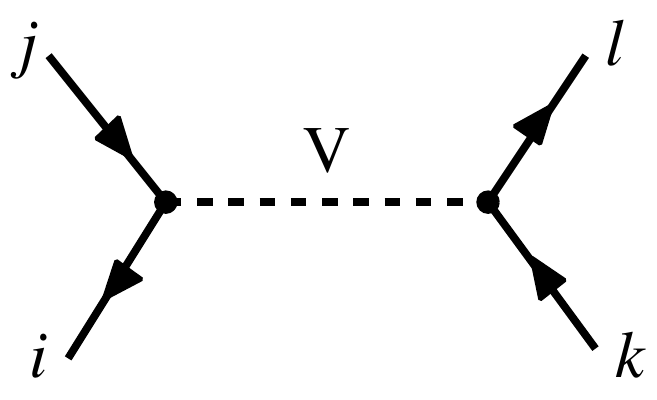}
\caption{Bare Coulomb interaction matrix element.}
\label{Fig5}
\end{figure}

The second step is to determine $L({\bf r}_1,{\bf r}_2;{\bf r}'_1,{\bf r}'_2,\omega)$ within the random phase approximation (RPA). We begin by inserting (\ref{KSgreen}) into (\ref{freefor}) to obtain the noninteracting 4-point polarizability: 
\begin{equation}
L_0({\bf r}_1,{\bf r}_2;{\bf r}'_1,{\bf r}'_2,\omega) =\sum_{ij}\frac{(f_j-f_i)\psi_i({\bf r}_1)\psi^*_j({\bf r}'_1)\psi_j({\bf r}_2)\psi^*_i({\bf r}'_2)}{\omega+\varepsilon_j-\varepsilon_i+i\eta\sgn(\varepsilon_i-\varepsilon_j)}\label{L0KS},
\end{equation}
where 
\begin{equation}
f_i=\left\{\begin{array}{cc}1;&i\le N\\0;&i>N\end{array}\right.
\end{equation} is the occupation factor and $N$ is the number of the occupied states. We can assume a similar expansion for $L$
\begin{equation}
L({\bf r}_1,{\bf r}_2;{\bf r}'_1,{\bf r}'_2,\omega)=\sum_{ijkl}\Theta_{ij}^{kl}
L^{kl}_{ij}(\omega)\psi_i({\bf r}_1)\psi^*_j({\bf r}'_1)\psi_l({\bf r}_2)\psi^*_k({\bf r}'_2),
\label{Lexpanz}
\end{equation}
where 
\begin{equation}
\Theta_{ij}^{kl} \equiv |f_j-f_i||f_l-f_k|,
\end{equation}
ensures that contributions to $L$ only come from transitions between empty and filled states.  That is, the summation indices should satisfy the following conditions: 
\begin{equation}
\begin{array}{llll}
i\le N,&j>N,&k\le N,&l>N\\
i\le N,&j>N,&k>N,&l\le N\\
i>N,&j\le N,&k\le N,&l>N\\
i>N,&j\le N,&k> N,&l\le N.
\end{array}
\end{equation}
After inserting (\ref{Lexpanz}) and (\ref{L0KS}) into (\ref{BSEomega}) and using the fact that in RPA the second term in the BSE kernel (\ref{BSEkerom}) (containing $W$) should be ignored, we obtain a matrix equation for $L$  
\begin{equation}
L^{kl}_{ij}(\omega)=L^{kl,0}_{ij}(\omega)+\sum_{i_1,j_1,k_1,l_1}\Theta_{ij}^{kl}L^{i_1j_1,0}_{ij}(\omega)\Xi^{k_1l_1,H}_{i_1j_1}L^{kl}_{k_1l_1}(\omega),
\label{mateqforL}
\end{equation}
where 
\begin{equation}
L^{kl,0}_{ij}(\omega)=\frac{f_j-f_i}{\omega+\varepsilon_j-\varepsilon_i+i\eta\sgn(\varepsilon_i-\varepsilon_j)}\delta_{ik}\delta_{jl}.
\label{njiki}
\end{equation}
The matrix of the BSE-Hartree kernel has the form
\begin{equation}
\Xi^{kl,H}_{ij}=2V^{kl}_{ij},
\label{BSEH}
\end{equation}
where the bare Coulomb interaction matrix elements
\begin{equation}
V^{kl}_{ij}=\int d{\bf r}_1d{\bf r}_2 \phi^j_i({\bf r}_1)V({\bf r}_1-{\bf r}_2)\phi^k_l({\bf r}_2),
\label{barclul}
\end{equation}
are shown in Fig.~\ref{Fig5}.
Here we introduced the two-particle wave functions 
\begin{equation}
\phi^j_i({\bf r})=\psi^*_i({\bf r})\psi_j({\bf r}),
\label{pairfun}
\end{equation}
and the factor of two in (\ref{BSEH}) comes from spin.

By solving equation (\ref{mateqforL}) we obtain matrix elements $L^{kl}_{ij}(\omega)$, and after inserting them into (\ref{Lexpanz}) we obtain the 4-point polarizability $L$ at the RPA level. Now the coordinate annihilation ${\bf r}_1={\bf r}'_1$ and ${\bf r}_2={\bf r}'_2$ gives us the response function (\ref{chi}) and by inserting it into (\ref{screened}) we obtain the propagator of the dynamically screened Coulomb interaction
\begin{eqnarray}
W({\bf r},{\bf r}',\omega)&=&V({\bf r},{\bf r}')+
\sum_{\alpha\beta\gamma\delta}\Theta_{\alpha\beta}^{\gamma\delta}
L^{\gamma\delta}_{\alpha\beta}(\omega)\nonumber\times\\
&&\int d{\bf r}_1d{\bf r}_2
V({\bf r},{\bf r}_1)\phi^\alpha_\beta ({\bf r}_1)
\phi^\delta_\gamma({\bf r}_2)
V({\bf r}_2,{\bf r}').
\label{Wexpanz}
\end{eqnarray}

The third step the is $G_0W_0$ approximation. After replacing the Green's function in (\ref{GWapprox}) by $G_0$ given by (\ref{KSgreen}), performing the $\omega$ integration in (\ref{GWapprox}), and using the spectral representation of the time ordered $W$, the exchange-correlation self energy becomes:\cite{GWtheory}
\begin{eqnarray}
\Sigma_{\textit{XC}}({\bf r},{\bf r}',\omega)&=&
\sum^{\infty}_{i=1}\psi_i({\bf r})\psi^*_i({\bf r}')\int^{\infty}_0d\omega'\frac{S({\bf r},{\bf r}',\omega')}{\omega-\varepsilon_i-\omega'+i\eta}
\nonumber\\
&&-\sum^{N}_{i=1}\psi_i({\bf r})\psi^*_i({\bf r}')W({\bf r},{\bf r}',\omega-\varepsilon_i).
\label{DinCOHSEX}
\end{eqnarray}
where the spectral function is defined as 
\begin{equation}
S({\bf r},{\bf r}',\omega)=-\frac{1}{\pi}\Imag\left\{ W({\bf r},{\bf r}',\omega)\right\}.
\label{spectralrep}
\end{equation}
Note that the Hartree term (\ref{Hartree}) is already included in the KS energies $\varepsilon_i$.

The first term in (\ref{DinCOHSEX}) is the screened exchange (SEX) term and the second term is the Coulomb hole correlation (COH) term, so this kind of $GW$ approximation is usually called the COHSEX approximation. After inserting (\ref{DinCOHSEX}) into a Dyson equation (\ref{dysonG}) we obtain a matrix equation for the Green's function 
\begin{equation}
G_{ij}(\omega)=G^0_{ij}(\omega)+G^0_{ik}(\omega)\Sigma^{\textit{XC}}_{kl}(\omega)G_{lj}(\omega),
\label{Deqm}
\end{equation}
where the self-energy matrix elements are defined by 
\begin{equation*}
\Sigma^{\textit{XC}}_{ij}=\left\langle \psi_i({\bf r})\left|\Sigma_{\textit{XC}}({\bf r},{\bf r}',\omega)\right|\psi_j({\bf r}')\right\rangle.
\end{equation*}
In most cases the off diagonal matrix elements of $\Sigma_{ij}$ weakly influence the diagonal matrix elements of $G_{ii}(\omega)$ \cite{GWtheory} and therefore we may safely neglect them. In this case (\ref{Deqm}) becomes a simple scalar equation and the solution is the quasiparticle Green's function
\begin{equation}
G_{ii}(\omega)\rightarrow G_{i}(\omega)=\frac{1}{\omega-(\varepsilon_i-V^{\textit{XC}}_i)-\Sigma^{\textit{XC}}_{i}(\omega)}.
\label{jukonja}
\end{equation}  
This represents the propagation of a quasiparticle (electron or hole) in state $i$.  The exchange-correlation self-energy is   likewise
\begin{equation}
\Sigma^{\textit{XC}}_{i}(\omega)=\left\langle \psi_i({\bf r})\left|\Sigma_{\textit{XC}}({\bf r},{\bf r}',\omega)\right|\psi_i({\bf r}')\right\rangle.
\label{sigmad}
\end{equation}
In (\ref{jukonja}) we had to subtract the KS exchange correlation contributions $V^{\textit{XC}}_i$ from the KS energy levels $\varepsilon_i$ because at this level of approximation these contributions are included in $\Sigma^{\textit{XC}}_{i}(\omega)$.
After inserting (\ref{DinCOHSEX}) into (\ref{sigmad}) and using expression (\ref{Wexpanz}), the exchange-correlation self energy becomes 
\begin{equation}
\Sigma^{\textit{XC}}_i(\omega)=
\sum^{\infty}_{j=1}\int^{\infty}_0d\omega' \frac{S_{ij}(\omega')}{\omega-\varepsilon_j-\omega'+i\eta}-\sum^{N}_{j=1}W_{ij}(\omega-\varepsilon_j)
\label{sigmafin}
\end{equation}
where 
\begin{equation}
W_{ij}(\omega)=V^{ij}_{ij}+\sum_{\alpha\beta\gamma\delta}V^{\alpha\beta}_{ij}L^{\gamma\delta}_{\alpha\beta}(\omega)V^{ij}_{\gamma\delta}
\label{Sigmamatr}
\end{equation}
and 
\begin{equation}
S_{ij}(\omega)=-\frac{1}{\pi}\Imag\left\{ W_{ij}(\omega)\right\}.
\label{speicka}
\end{equation}
The first term in (\ref{sigmafin}) represents the bare and induced exchange (Fock) energy and the second term represents the Coulomb hole correlation energy. This corresponds to the polarization energy shift due to an extra electron or hole in the system. 

It should be noted here that even though for unoccupied states the KS-Hartree term is exact, for occupied states it contains an incorrect self-interaction term. However, the self-energy $\Sigma^{\textit{XC}}_{i}(\omega)$ also contains an incorrect self-interaction Fock term $V^{ii}_{ii}$, equal in amount and with opposite sign, so that these two terms exactly cancel. Therefore, the self-interaction Fock term $V^{ii}_{ii}$ is useful and should not be extracted from $\Sigma^{\textit{XC}}_{i}(\omega)$.

The poles of equation (\ref{jukonja}) 
\begin{equation}
\omega-(\varepsilon_i-V^{\textit{XC}}_i)-\Sigma^{\textit{XC}}_{i}(\omega)=0
\label{qpEnergyeq}
\end{equation}
represent the new quasiparticle energies $\varepsilon^{\textit{QP}}_{i}$. In the quasiparticle approach the solution of (\ref{qpEnergyeq}) is close to the real axis and we can expand equation 
(\ref{qpEnergyeq}) around $\varepsilon_i$  as
\begin{equation}
V^{\textit{XC}}_i-\Sigma^{\textit{XC}}_{i}(\varepsilon_i)+\left[1-\left.\frac{\partial\Sigma^{\textit{XC}}_{i}(\omega)}{\partial\omega}\right|_{\varepsilon_i}\right](\omega-\varepsilon_i)=0     ,
\end{equation}
and the solution is
\begin{equation}
\varepsilon^{\textit{QP}}_{i}=\varepsilon_i+Z_i \left[\Sigma^{\textit{XC}}_{i}(\varepsilon_i)-V^{\textit{XC}}_i\right],
\label{QPenergy}
\end{equation}
where we introduced the normalization factor
\begin{equation}
Z_i=\left[1-\left.\frac{\partial \Sigma^{\textit{XC}}_{i}(\omega)}{\partial\omega}\right|_{\varepsilon_i}\right]^{-1}.
\label{reziduum}
\end{equation}
In the quasiparticle approach the imaginary part of $\Sigma^{\textit{XC}}_{i}$ can be neglected and the new, renormalized quasiparticle Green's function becomes
\begin{equation}
G_{\textit{QP}}({\bf r},{\bf r}',\omega)=\sum_i\frac{\psi_i({\bf r})\psi^*_i({\bf r}')}{\omega-\varepsilon^{\textit{QP}}_i+i\eta\sgn(\varepsilon_F-\varepsilon^{\textit{QP}}_i)}.
\label{KSQP}
\end{equation}

\begin{figure}[h]
\centering
\includegraphics[width=\columnwidth]{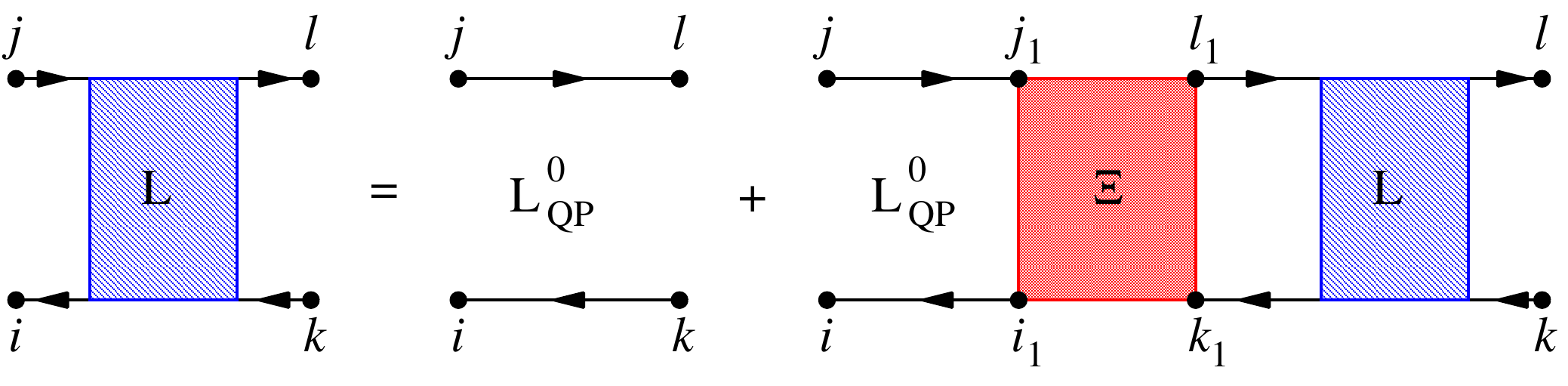}
\caption{(Color online) Bethe-Salpeter equation.}
\label{Fig6}
\end{figure}

The fourth and final step is solving the full Bethe-Salpeter equation (\ref{BSEomega}). First we insert the quasiparticle Green's functions (\ref{KSQP}) into (\ref{freefor}) to obtain the noninteracting quasiparticle 4-point polarizability $L^0_{\textit{QP}}$. This has the same form as (\ref{L0KS}) except that $\varepsilon_i$ is replaced by $\varepsilon^{\textit{QP}}_i$. Then, by using $L^0_{\textit{QP}}$ and repeating the RPA scheme (\ref{Lexpanz}--\ref{Wexpanz}) we obtain a new 4-point polarizability $L_{\textit{QP}}$ and dynamically screened Coulomb propagator $W$. Now we can solve the Bethe-Salpeter equation (\ref{BSEomega}) with the full kernel (\ref{BSEkerom}). After inserting $L^0_{\textit{QP}}$, $W(\omega=0)$, and the expansion (\ref{Lexpanz}) into (\ref{BSEomega}) and (\ref{BSEkerom}) the BS matrix equation becomes  
\begin{equation}
L^{kl}_{ij}(\omega)=L^{kl,0}_{ij\ \textit{QP}}(\omega)+
\sum_{i_1j_1k_1l_1}\Theta_{i_1 j_1}^{k_1 l_1}L^{i_1j_1,0}_{ij\ \textit{QP}}(\omega) \Xi^{k_1l_1}_{i_1j_1} L^{kl}_{k_1l_1}(\omega),
\label{mateqforBSE}
\end{equation}
as shown in Feynman diagrams in Fig.~\ref{Fig6}.
The matrix of noninteracting quasiparticle 4-point polarizability has the form  
\begin{equation}
L^{kl,0}_{ij\ \textit{QP}}(\omega)=
\frac{f_j-f_i}{\omega+\varepsilon^{\textit{QP}}_j-\varepsilon^{\textit{QP}}_i+i\eta\sgn(\varepsilon^{\textit{QP}}_i-\varepsilon^{\textit{QP}}_j)}\delta_{ik}\delta_{jl},
\label{iqppol}
\end{equation}
and the BSE kernel consists of two terms 
\begin{equation}
\Xi^{kl}_{ij}=\Xi^{kl,H}_{ij}-\Xi^{kl,F}_{ij}.
\label{barculkul}
\end{equation}
This approximation is usually called the Time Dependent Screened Hartree Fock Approximation (TDSHFA). The first term in (\ref{barculkul}) is a matrix of the BSE-Hartree kernel given by (\ref{BSEH}) and (\ref{barclul}), and the second term is a matrix of the BSE-Fock kernel given by 
\begin{equation}
\Xi^{kl,F}_{ij}=\int d{\bf r}_1d{\bf r}_2 \phi^j_l({\bf r}_1)W_{\textit{QP}}({\bf r}_1,{\bf r}_2,\omega=0)\phi^k_i({\bf r}_2).
\end{equation}

After using (\ref{Wexpanz}) the BSE-Fock kernel can be expressed in terms of the 4-point polarizability matrix as
$L_{\textit{QP}}$  
\begin{equation}
\Xi^{kl,F}_{ij}=V^{ki}_{lj}+\sum_{\alpha\beta\gamma\delta}\Theta_{\alpha\beta}^{\gamma\delta}V^{\alpha\beta}_{lj} L^{\gamma\delta}_{\alpha\beta\ \textit{QP}}(\omega=0) V^{ki}_{\gamma\delta}.
\label{BSEFOCK}
\end{equation}
the Fock term does not contain the spin factor $2$ because it does not allow a spin flip.  This will be discussed later in more detail.

\begin{figure}[h]
\centering
\includegraphics[width=\columnwidth]{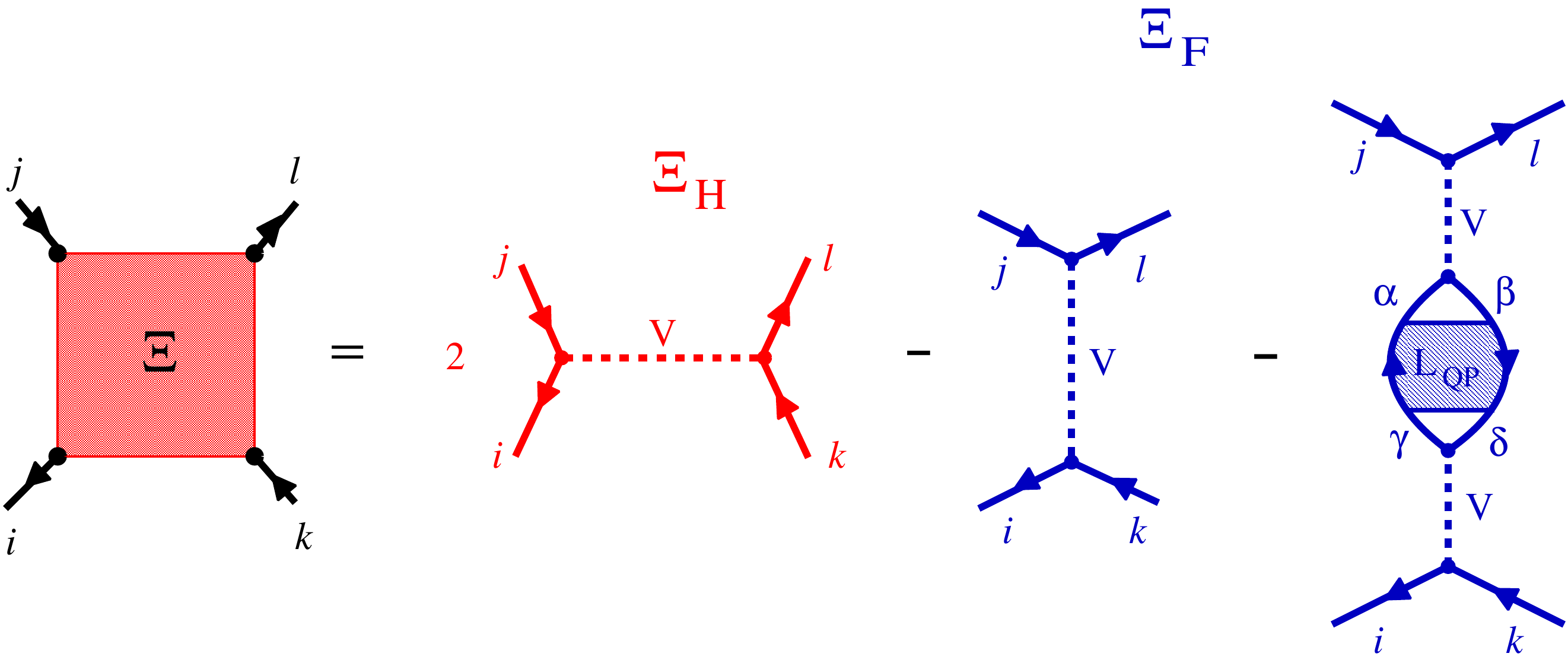}
\caption{(Color online) BSE kernel in TDSHF approximation. First term represents the BSE-Hartree kernel, the second term 
represents the bare BSE-Fock kernel, and the third term represents the induced BSE-Fock kernel.}
\label{Fig7}
\end{figure}

The total BSE kernel is shown in Feynman diagrams in Fig.~\ref{Fig7}. The first term is the BSE-Hartree or RPA term.  This represents the Coulomb interaction between electron-hole creation and annihilation.  The second term is the BSE-Fock term.  This represents the interaction between electron and hole mediated by the bare Coulomb interaction.  The third term is the induced BSE-Fock term.  This represents the interaction between electron and hole mediated by 
the induced Coulomb interaction. The second and the third term together is also called the screened BSE-Fock term.

After solving equation (\ref{mateqforBSE}) by using expansion (\ref{Lexpanz}) and coordinate annihilation ${\bf r}_1={\bf r}_1'$ and ${\bf r}_2={\bf r}_2'$, we obtain the BS response function 
\begin{eqnarray}
\chi({\bf r}_1,{\bf r}_2,\omega)&=&L({\bf r}_1,{\bf r}_2;{\bf r}_1,{\bf r}_2,\omega)
\nonumber\\
&=&\sum_{ijkl}\Theta_{ij}^{kl}L^{kl}_{ij}(\omega)\psi_i({\bf r}_1)\psi^*_j({\bf r}_1)\psi_l({\bf r}_2)\psi^*_k({\bf r}_2).
\label{Chi2point}
\end{eqnarray}

\subsection{Optical absorption spectra}
\label{optabs}

Optical absorption spectra is usually calculated by solving the BSE in the resonant approximation. This means that the frequency dependent part of (\ref{Lexpanz}), i.e.\ $L^{kl}_{ij}(\omega)$, is represented as a sum over harmonic oscillators with frequencies $\omega_S$ representing the frequencies of the excitations in system, and amplitudes $A_S$ representing their oscillator strengths. This reduces the BS equation (\ref{BSEomega}) into a standard eigenvalue problem for $\omega_S$ and $A_S$.  This can then be inserted into the absorption spectra formula.\cite{StevenLouie2000,Rubio} In this subsection we shall describe an alternative approach in which the optical absorption spectra is obtained directly from  $L^{kl}_{ij}(\omega)$.

In an optical absorption experiment the incident electromagnetic wave couples to the electronic excitations in the system and is partially absorbed. In linear response theory the power at 
which the external electromagnetic energy is absorbed in the system can be obtained from the expression 
\begin{equation}
P(t)=\int^{\infty}_{-\infty}dt_1{\bf E}^{\textit{ext}}({\bf r}_1,t)\otimes_1\Pi({\bf r}_1,{\bf r}_2,t-t_1)\otimes_2{\bf A}^{\textit{ext}}({\bf r}_2,t_1),
\label{powerloss}
\end{equation}
where $\Pi$ is the current-current response function of the system, while ${\bf E}^{\textit{ext}}$ and ${\bf A}^{\textit{ext}}$ are external electric field and vector potential respectively. The symbol $\otimes$ denotes convolution in real space.

We shall assume that the incident electromagnetic field is a plane wave of unit amplitude
\begin{equation}
{\bf A}^{\textit{ext}}({\bf r},t)={\bf e}\cos({\bf k}{\bf r}-\omega t), 
\label{incidentemp}
\end{equation}
where ${\bf e}$ is the polarization vector. If we also assume that the external scalar potential is $\Phi^{\textit{ext}}=0$, this implies that ${\bf E}^{\textit{ext}}=-\frac{1}{c}\frac{\partial{\bf A}^{\textit{ext}}}{\partial t}$. If the wavelength ${\lambda}$ is much larger than the dimension of the illuminated system or the crystal unit cell, the dipole approximation can be applied and the absorption power becomes
\begin{equation}
P(\omega)=-\omega \Imag\left\{\sum_{\mu\nu}e_\mu e_\nu\int d{\bf r}_1{\bf r}_2\Pi_{\mu\nu}({\bf r}_1,{\bf r}_2,\omega)\right\}. 
\label{apspower}
\end{equation}

In the Coulomb gauge ($\nabla\dotproduct {\bf A}=0$), there is an instantaneous interaction mediated by the Coulomb interaction $V$ and a transversal interaction that is retarded and mediated by photons. In small systems such as a molecule, the interaction between charge/current fluctuations mediated by photons is negligible compared to the Coulomb interaction. This allows us to describe all interactions inside the molecule by the instantaneous Coulomb interaction $V$ and the interaction of the molecule with the environment by both interactions.  In this case this is only interactions with photons described by ${\bf A}^{\textit{ext}}$.

\begin{figure}[ht]
\centering
\includegraphics[width=\columnwidth]{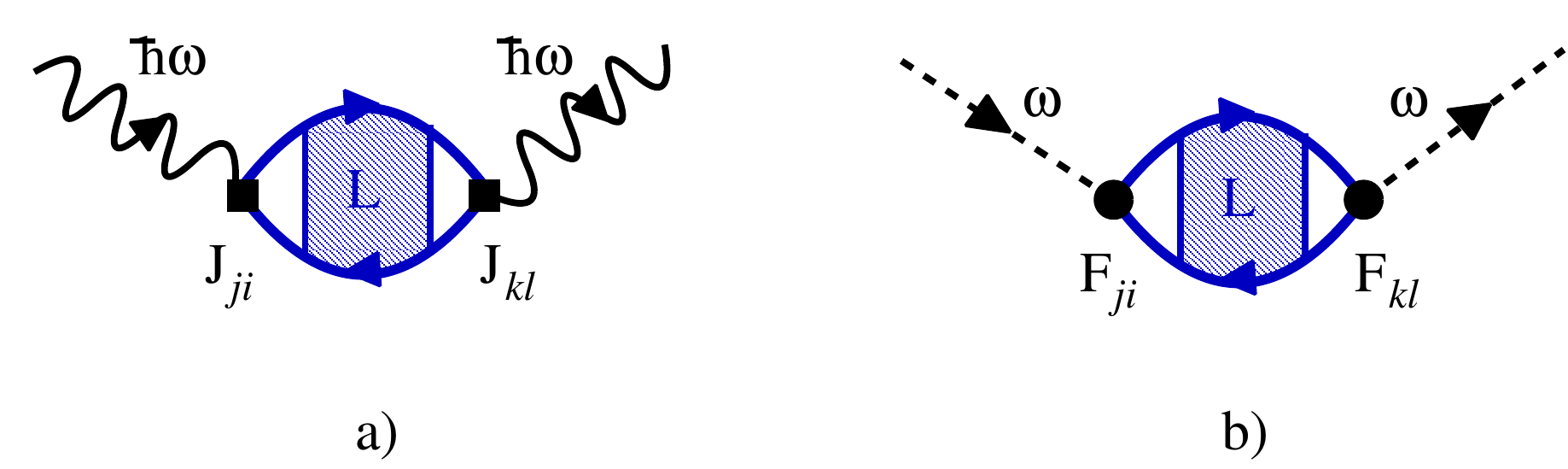}
\caption{(Color online) Feynman diagrams for a) optical absorption process and b) energy loss process. The squares represent current vertices and dots charge vertices.}
\label{Fig8}
\end{figure}

 As a result, the current-current response function can be 
expressed in terms of the response function (\ref{Chi2point}), except that now charge vertices should be replaced by current vertices (shown as squares in Fig.~\ref{Fig8}) to get   
\begin{equation}
\Pi_{\mu\nu}({\bf r},{\bf r}',\omega)=
\frac{e^2\hslash}{m^2c}\sum_{ijkl}\Theta_{ij}^{kl}L^{kl}_{ij}(\omega)\psi^*_j({\bf r})\nabla_\mu\psi_i({\bf r})\psi^*_k({\bf r}')\nabla_\nu\psi_l({\bf r}').
\label{Piexpanz}
\end{equation}

After inserting (\ref{Piexpanz}) into (\ref{apspower}) the absorption power becomes: 
\begin{equation}
P(\omega)=-\omega \Imag\left\{\sum_{ijkl}\Theta_{ij}^{kl}L^{kl}_{ij}(\omega)J_{ji}J_{kl}\right\}
\label{apspowerF}
\end{equation}
where the form factors are
\begin{equation}
J_{ij}=\sum_{\mu}e_\mu\int d{\bf r} \psi^*_i({\bf r})\nabla_\mu\psi_j({\bf r}).
\label{Formfact}
\end{equation}

\subsection{Energy loss spectra}
\label{abspowerr}

Since some of the excitation modes, so-called dark modes, cannot be excited with incident electromagnetic waves, and here we are interested in all types of excitations in the molecule, we have to design some alternative probe which is able to excite the dark modes as well. The simplest choice is an external time dependent charge distribution. The rate at which an external charge distribution is losing energy to excitations in the system is given by \cite{wake}   
\begin{equation}
P(t)=\int^{\infty}_{-\infty}dt_1 \Phi^{\textit{ext}}({\bf r}_1,t)\otimes_1\chi({\bf r}_1,{\bf r}_2,t-t_1)\otimes_2\Phi^{\textit{ext}}({\bf r}_2,t_1) 
\label{energyloss}
\end{equation}
where $\Phi^{\textit{ext}}({\bf r},t)$ is the time dependent potential produced by the external charge, $\chi$ is the density-density response function of the system, and $\otimes$ denotes convolution in the real space. If we assume a simple oscillatory time dependence
\begin{equation}
\Phi^{\textit{ext}}({\bf r},t)=\Phi^{\textit{ext}}({\bf r})\cos\omega t
\end{equation}
the power loss becomes: 
\begin{equation}
P(\omega)=-\omega \Imag\left\{\int d{\bf r}_1{\bf r}_2 \Phi^{\textit{ext}}({\bf r}_1)\chi({\bf r}_1,{\bf r}_2,\omega)\Phi^{\textit{ext}}({\bf r}_2)\right\}. 
\label{energyloss1}
\end{equation}
We note that equation (\ref{energyloss1}) is the longitudinal equivalent of equation (\ref{apspower}). After using the expression for the response function 
(\ref{Chi2point}) and definition 
(\ref{pairfun}) the power loss can also be written in term of matrix elements $L^{kl}_{ij}(\omega)$ as
\begin{equation}
P(\omega)=-\omega \Imag\left\{\sum_{ijkl}\Theta_{ij}^{kl}L^{kl}_{ij}(\omega)F^*_{ij}F_{kl}\right\},
\label{energyloss2}
\end{equation}
where the form factors are 
\begin{equation}
F_{ij}=\int d{\bf r}\phi^j_i({\bf r})\Phi^{\textit{ext}}({\bf r}). 
\label{Floss}
\end{equation}

\section{Application to molecular spectroscopy}
\label{Apptomol}

In this section we shall show how we may apply the general procedure described in Sec.~\ref{generaltheo} to calculate the optical absorption and energy loss spectra of benzene in gas phase, deposited on graphene, and adsorbed on a metallic substrate.

\subsection{Spectroscopy of gaseous benzene} 
\label{benzene-izolated}

\subsubsection{Numerical solution of the BSE}

The first step is to determine the molecular ground state electronic structure. The benzene Kohn-Sham orbitals $\psi_i({\bf r})$ and energy levels $\varepsilon_i$ are obtained by using the plane-wave self-consistent field density functional theory (DFT) code (PWscf), within the Quantum Espresso (QE) package,\cite{QE} using the Perdew-Wang GGA (PW91) exchange and correlation (xc)-functional.\cite{PW91-GGA} We model the benzene molecule using a periodically repeated $22.845\ a_0 \times 22.845\ a_0\times 22.845\ a_0$ unit cell. Since there is no intermolecular overlap, the ground state electronic density is calculated at the $\Gamma$ point only. For carbon and hydrogen atoms we used GGA-based ultra soft pseudo potentials,\cite{pseudopotentials} and found the energy spectrum to be convergent with a $30$~Ry plane-wave cutoff.

 The benzene molecule has $30$ valence electrons, which corresponds to $15$ doubly occupied valence orbitals. For the 4-point polarizability calculation we use $60$ orbitals, i.e.\ $15$ occupied and $45$ unoccupied orbitals. In Sec.~\ref{Results} we will show that the excitation spectrum is mostly defined by transitions inside the $\pi-\pi^*$ complex, or between occupied states $a_{2u},e_{1g},e_{1g}$ and unoccupied states $e_{2u},e_{2u},b_{2g}$.\cite{Spectro3}
 
The KS wave functions are periodic and can be Fourier expanded as 
\begin{equation}
\Psi_{i}({\bf r})=\frac{1}{\sqrt{\Omega}}\sum_{\bf G}C_{i}({\bf G})e^{i{\bf G}{\bf r}},
\label{KSexp}
\end{equation}
where ${\bf G}$ are reciprocal vectors and $\Omega$ is the normalization volume. In this case the two particle wave functions defined by (\ref{pairfun}) are also periodic and can be expanded as  
\begin{equation}
\phi^j_i({\bf r})=\sum_{\bf G}C^j_i({\bf G})e^{i{\bf G}\dotproduct{\bf r}},
\label{pairfunexp}
\end{equation}
where the Fourier coefficients $C^j_i({\bf G})=\frac{1}{\Omega}\int d{\bf r}e^{-i{\bf G}\dotproduct{\bf r}}\phi^j_i({\bf r})$, with use of expansion (\ref{KSexp}) and definition (\ref{pairfun}), can be expressed in terms of coefficients $C_{i}({\bf G})$ as 
\begin{equation}
C^j_i({\bf G})=\frac{1}{\Omega}\sum_{{\bf G}_1}C^*_{i}({\bf G}_1)C_{j}({\bf G}_1+{\bf G}).
\label{Furcoef}
\end{equation}
This transformation enables higher numerical efficiency in the calculation of the bare Coulomb interaction matrix elements $V^{kl}_{ij}$ which are the most frequently used quantities throughout the calculation.   

Since we study a single isolated benzene molecule, we have to exclude the effect on its polarizability due to the interaction with surrounding molecules in the lattice. This is accomplished using the truncated Coulomb interaction \cite{radcutof}
\begin{equation} 
V_C({\bf r}-{\bf r}')=\frac{\Theta\left(|{\bf r}-{\bf r}'|-R_C\right)}{|{\bf r}-{\bf r}'|},
\label{radcutof}
\end{equation}
where $\Theta$ is the Heaviside step function, and $R_C$ is the range of the Coulomb interactions, i.e. the radial cutoff. Since we choose the lattice constant $L = 22.845\ a_0$ to be more then twice the range of the benzene molecule's density, choosing the radial cutoff to be $R_C=L/2$ ensures that the charge fluctuations created within the molecule produce a field throughout the whole molecule but do not produce any field within the surrounding molecules. The definition (\ref{radcutof}) is very useful because the Coulomb interaction remains translationally invariant. This leads to a simple Fourier transform
\begin{equation}
V_C(q)=\frac{4\pi}{q^2}[1-\cos qR_C].
\label{ftci}
\end{equation}
After using the definition (\ref{barclul}), the Fourier transform of the Coulomb interaction (\ref{ftci}) and the expansion (\ref{pairfunexp}), the bare Coulomb interaction matrix elements become: 
\begin{equation}
V^{kl}_{ij}=\frac{1}{\Omega_{\text{cell}}}\sum_{{\bf G}}C^j_i({\bf G})[C^l_{k}({\bf G})]^*V_C({\bf G}),
\label{barcoulnes}
\end{equation}
where $\Omega_{\text{cell}}=L^3$ is the unit cell volume. 

After we obtain the KS spectra $\varepsilon_i$ and Coulomb matrix elements (\ref{barcoulnes}) we can perform the RPA scheme (\ref{mateqforL}--\ref{BSEH}). The matrix of the noninteracting 4-point polarizability $L^{kl,0}_{ij}$  (\ref{njiki}) can be obtained directly from the KS energies $\varepsilon_i$, and the BSE-Hartree kernel (\ref{BSEH}) directly from the matrix elements (\ref{barcoulnes}). This gives us the 4-point polarizability matrix $L^{kl}_{ij}(\omega)$, which is needed to obtain $W_{ij}(\omega)$ through (\ref{Sigmamatr}) and finally the exchange-correlation self energy $\Sigma^{\textit{XC}}_i(\omega)$ using (\ref{sigmafin}). From $\Sigma^{\textit{XC}}_i(\omega)$ we now obtain corrected quasiparticle energies $\varepsilon^{\textit{QP}}_i$ from (\ref{QPenergy}) and the matrix of noninteracting quasiparticle 4-point polarizability $L^{kl,0}_{ij\ \textit{QP}}(\omega)$ from (\ref{iqppol}). Using $L^{kl,0}_{ij\ \textit{QP}}(\omega)$ and repeating the RPA scheme (\ref{Lexpanz}--\ref{Wexpanz}) we obtain a new 4-point polarizability $L^{kl}_{ij\ \textit{QP}}(\omega)$ and BSE-Fock kernel given by (\ref{BSEFOCK}). Finally,  using $L^{kl,0}_{ij\ \textit{QP}}(\omega)$, the BSE-Hartree kernel (\ref{BSEH}), and the BSE-Fock kernel (\ref{BSEFOCK}) we can solve the Bethe-Salpeter matrix equation (\ref{mateqforBSE}) for $L^{kl}_{ij}(\omega)$. 

The fact that we use $15$ occupied and $45$ unoccupied orbitals for the calculation means that the dimension of the Bethe-Salpeter kernel matrix is $1350\times 1350$.  However, this does not depend on the number of plane waves used in the expansion of the Coulomb interaction matrix elements (\ref{barcoulnes}), which is an important advantage of our method. This accelerates matrix calculations and at the same time allows us to perform very accurate calculation of the Coulomb matrix elements. For example, in expression (\ref{barcoulnes}) we use a 30~Ry energy cut off, which corresponds to an expansion over $35000$ plane waves. The disadvantage of this method is that the dimension of the matrix $V^{kl}_{ij}$ increases with the number of occupied states $N$.  This means the method is not computationally efficient for very large molecules.

\subsubsection{Determination of the energy loss and optical absorption spectra}

\begin{figure}
\centering
\includegraphics[width=\columnwidth]{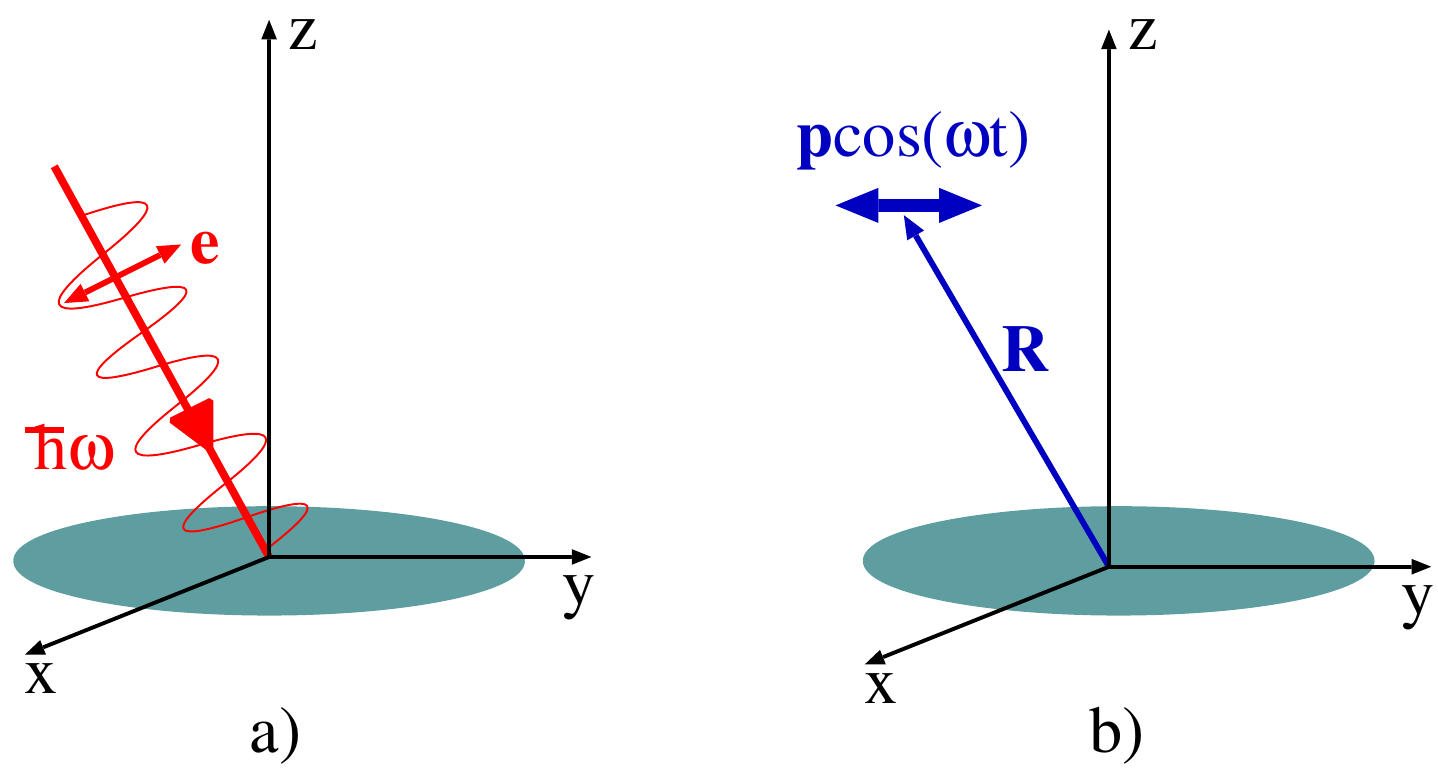}
\caption{(Color online) Schematic representation of an a) optical absorption experiment and a b) dipole energy loss experiment.}
\label{Fig9}
\end{figure}

To be able to detect all types of electronic modes in the molecule, we simulate two kind of experimental spectroscopic methods. First is an optical absorption experiment.  This is simulated by the absorption of a plane wave of light illuminating the molecule, as shown schematically in Fig.~\ref{Fig9}a). Second is an energy loss experiment.  This is simulated by the energy loss of an oscillating dipole placed close the molecule, as shown schematically in Fig.~\ref{Fig9}b). 

The benzene absorption spectra is obtained using expression (\ref{apspowerF}), where the form factors (\ref{Formfact}) after using the expansion (\ref{KSexp}) become 
\begin{equation}
J_{ij}=i\sum_{{\bf G}}[{\bf e}\dotproduct{\bf G}]C^*_i({\bf G})C_j({\bf G}).
\label{currf}
\end{equation}
Electronic modes are also excited by an external potential $\Phi^{\textit{ext}}(t)$ where the rate at which the probe is losing energy is given by expressions (\ref{energyloss2}) and (\ref{Floss}). After inserting the Fourier transform of the external potential 
\begin{equation}
\Phi^{\textit{ext}}({\bf r})=\int \frac{d{\bf q}}{(2\pi)^3}\Phi^{\textit{ext}}({\bf q})e^{i{\bf q}\dotproduct{\bf r}}  
\label{Furtrans}
\end{equation}
in the definition of the form factors (\ref{Floss}), using the Fourier expansion (\ref{pairfunexp}), and the fact that the $\Phi^{\textit{ext}}({\bf r})$ is a real function, i.e.\ $\Phi^{\textit{ext}}(-{\bf q})=[\Phi^{\textit{ext}}({\bf q})]^*$, the form factors become: 
\begin{equation}
F_{ij}=\sum_{\bf G}C^j_i({\bf G})[\Phi^{\textit{ext}}({\bf G})]^*. 
\label{Flossnew}
\end{equation}
We model the external probe as a dipole with dipole moment ${\bf p}$ placed at position ${\bf R}$ from the center of the benzene molecule, as shown in Fig.~\ref{Fig9}b). In this case, the 
Fourier transform of the external potential has the explicit form          
\begin{equation}
\Phi^{\textit{ext}}({\bf q})=-i\frac{4\pi}{q^2}e^{-i{\bf q}\dotproduct{\bf R}}{\bf q}\dotproduct{\bf p}.  
\label{Ftcoef}
\end{equation}

\subsubsection{Singlet and triplet excitons}

Besides the spatial symmetry of the molecular electronic excitations, which determines whether the excitation will be dark or bright, there are also two classes of solution of the BSE with respect to spin. If the spin-orbit interaction is negligible compared with the electron-hole interaction, as we assume here, then each quasiparticle state has an additional quantum number associated with spin, i.e.\ up $\uparrow$ or down $\downarrow$. This has a simple impact on BSE. If spins of an excited electron-hole pair are parallel (e.g.\ spins of states $i$ and $j$ in Fig.~\ref{Fig7} are both $\uparrow$) then the final state Hartree interaction can either leave the spin configuration unchanged or flip both spins into the opposite direction (i.e.\ spins of states $k$ and $l$ are both $\downarrow$).  However, they will remain parallel.  On the other hand, the Fock interaction always leaves the spin configuration unchanged, and this is why there is no factor of two in the Fock kernel (\ref{BSEFOCK}). The BSE kernel is then simply $\Xi=\Xi^H-\Xi^F$, and any excitons created in this way have a spin singlet configuration. 

If the external perturbation instead creates an electron-hole pair with anti-parallel spins  (e.g.\ spins of states $i$ and $j$ 
are $\uparrow$ and $\downarrow$ respectively) then, because of the orthogonality, such a pair cannot be annihilated and the Hartree interaction is inactive. The Fock interaction, responsible for the mutual electron-electron and hole-hole scattering, survives and it does not change the initial spin configuration, i.e.\ the final spins are still anti-parallel. The BSE kernel then consists of the Fock term only, i.e.\ $\Xi=-\Xi^F$ and this type of exciton forms a spin triplet configuration.\cite{StevenLouie2000} Both spin classes of excitons will be investigated in Sec.~\ref{spectracopyofbenzeneongraphene}.

\subsection{Spectroscopy of the benzene deposited on graphene} 
 \label{spectracopyofbenzeneongraphene}

\begin{figure}[h]
\centering
\includegraphics[width=\columnwidth]{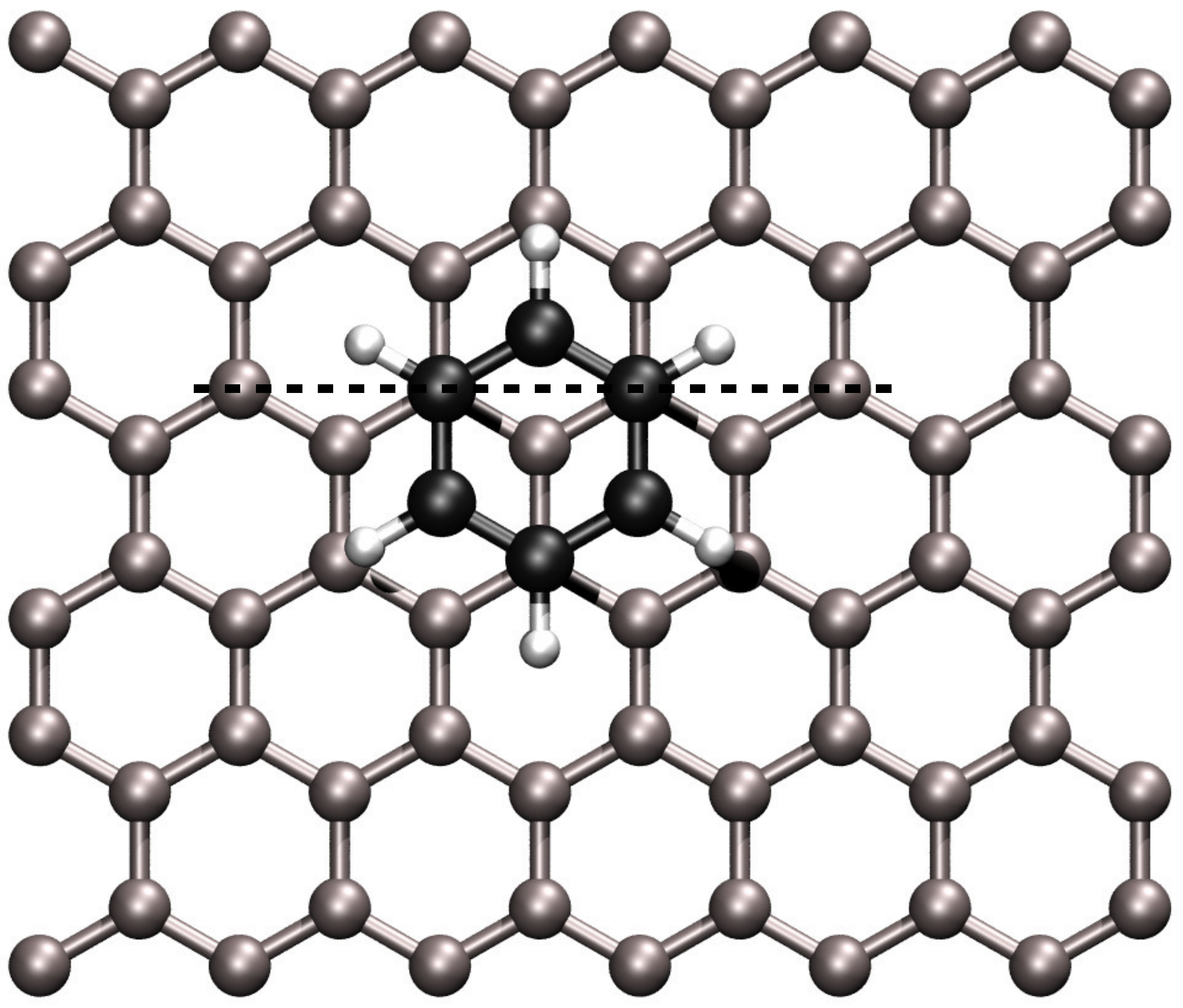}
\caption{(Color online) Ground state geometry of a benzene molecule deposited on graphene. Benzene carbon and hydrogen atoms are depicted as black and white spheres respectively, while carbon atoms of the graphene substrate are depicted as gray spheres.}
\label{Fig10}
\end{figure}

We next investigate the energy levels, optical absorption and energy loss spectra of benzene deposited on a graphene substrate, as illustrated in Fig.~\ref{Fig10}. 

\begin{figure}
\centering
\includegraphics[width=\columnwidth]{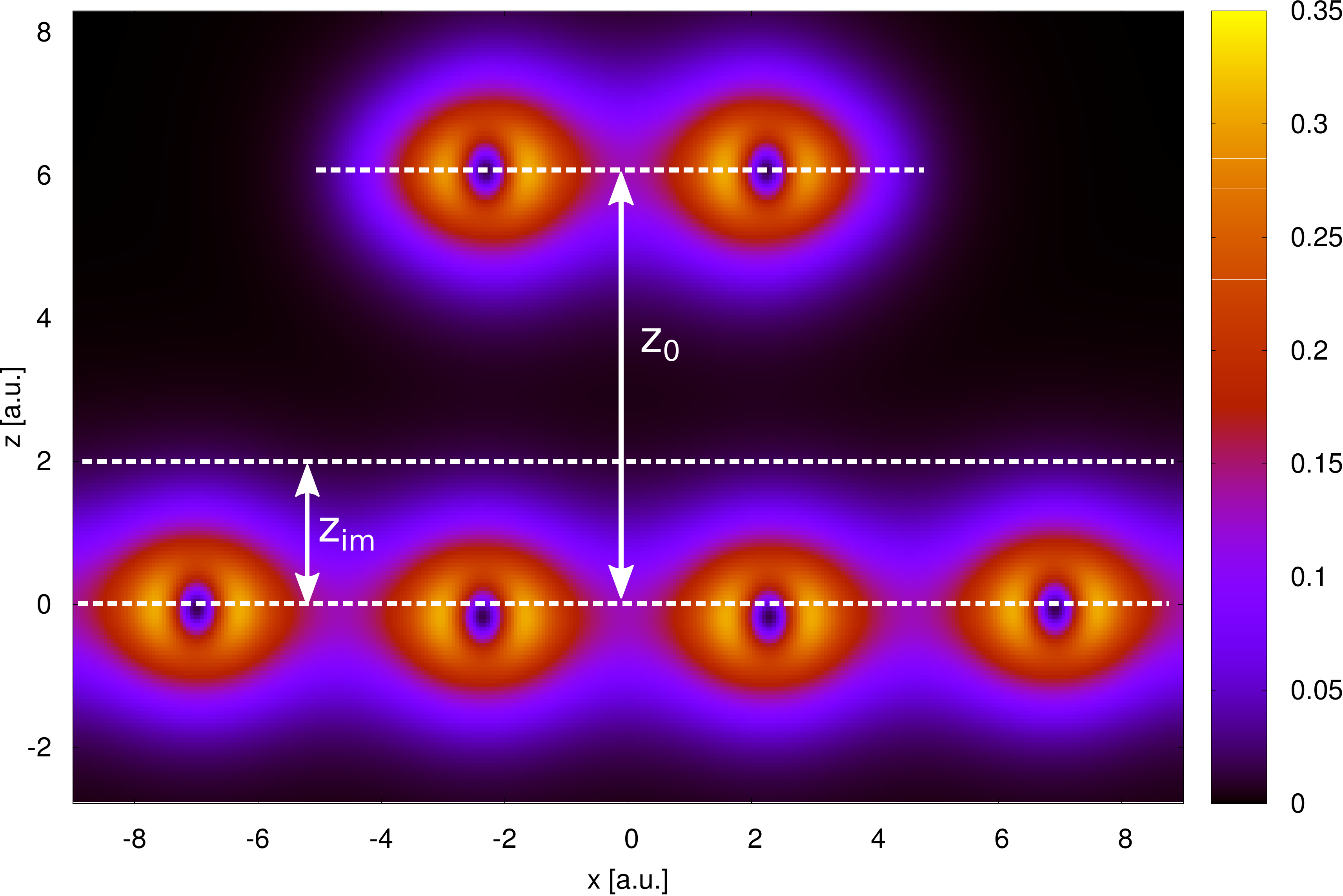}
\caption{(Color online) Ground state electronic density of benzene horizontally deposited on graphene. Position of the graphene image plane is denoted by $z_{im}$ and the equilibrium benzene-graphene separation by $z_0$. The density is plotted in the
$xz$-plane across the dashed line denoted in Fig.~\ref{Fig10}.}
\label{Fig11}
\end{figure}

Graphene and benzene planes are chosen to lie parallel to the $xy$-plane, i.e.\ have a normal parallel to the $z$-axis. Ground state electronic and crystal structure is obtained by structural relaxation using the second version of the van der Waals density functional of Lee \emph{et al.}\cite{lee} and the exchange functional (C09) developed by Cooper.\cite{cooper} This  combination of functionals gives good agreement with experimental data for similar systems.\cite{hamada} The initial, most favourable, geometry is taken from Ref.~\onlinecite{benzongra}. For obtaining ground state electronic density we used a supercell with dimensions $a\times \frac{\sqrt{3}}{2}a \times a$ where $a=27.906\ a_0$. We employed a plane wave basis set with ultrasoft pseudopotentials as implemented in QE.\cite{QE,pseudopotentials} The kinetic energy cutoff for the plane waves was $40$ Ry, and $500$~Ry for the density. We applied an $8\times 8 \times 1$ Monkhorst-Pack special \textbf{k}-point mesh to sample the Brillouin Zone. For the average equilibrium separation between benzene and graphene we obtain $z_{0} \approx 6\ a_0$. This is the same as that reported in Ref.~\onlinecite{benzongra}. Because of the large separation, the electronic densities of these systems do not overlap, as can be clearly seen in Fig.~\ref{Fig11}.  This shows the ground state electronic density in the $xz$-plane along the dashed line denoted in Fig.~\ref{Fig10}.

\begin{figure}[h]
\centering
\includegraphics[width=\columnwidth]{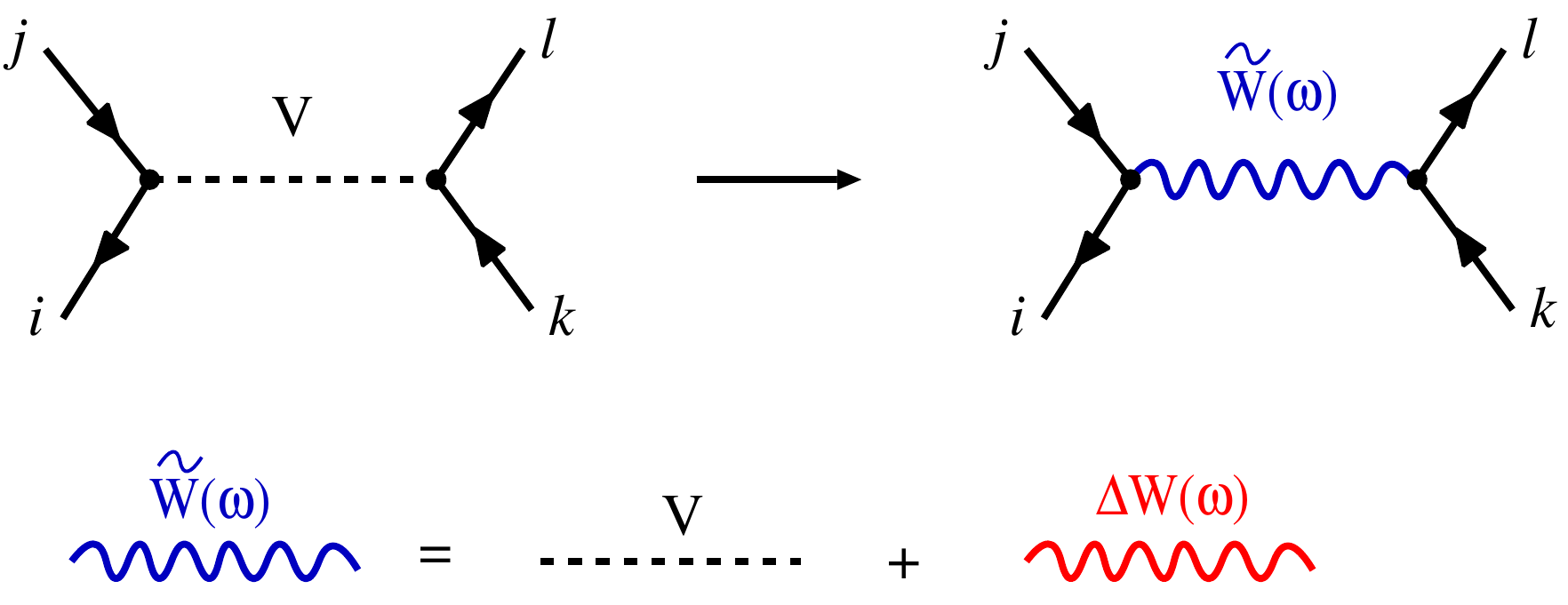}
\caption{(Color online) Screening of the bare Coulomb interaction by the polarization of the substrate.}
\label{Fig12}
\end{figure}

 The fact that the electronic densities do not overlap simplifies the impact of the graphene to benzene energy spectra and response function significantly. More specifically, since there is no inter-system electron hopping, the only modification comes from the additional screening caused by polarization of the graphene.  In other words, interactions between charge fluctuations in the benzene have to be additionally screened because of the polarization of 
the graphene. This simply means that the bare Coulomb interaction has to be modified as follows:
\begin{equation}
V({\bf r},{\bf r}')\rightarrow\tilde{W}({\bf r},{\bf r}',\omega)=V({\bf r},{\bf r}',\omega)+\Delta W({\bf r},{\bf r}',\omega),
\label{indc}
\end{equation}
and also shown in Feynman diagrams in Fig.~\ref{Fig12}.  Here $\Delta W$ is the induced dynamically screened Coulomb interaction of the graphene.\cite{Duncan2} Consequently, the matrix elements (\ref{barclul}) have to be modified to    
\begin{equation}
V^{kl}_{ij}\rightarrow \tilde{V}^{kl}_{ij}(\omega)=V^{kl}_{ij}+\Delta W^{kl}_{ij}(\omega),
\label{indcmatr}
\end{equation}
where 
\begin{equation}
\Delta{W}^{kl}_{ij}(\omega)=\int_{\Omega_{\text{cell}}}d{\bf r}_1d{\bf r}_2 \phi^j_i({\bf r}_1)\Delta{W}({\bf r}_1,{\bf r}_2,\omega)\phi^k_l({\bf r}_2).
\label{indmatrel}
\end{equation}

We note that the integration in (\ref{indmatrel}) is performed over the benzene superlattice unit cell volume $\Omega_{\text{cell}}=L^3$. By doing this we avoid the graphene mediated intermolecular interaction, i.e.\ the influence of the surrounding molecules through polarization of the graphene is completely excluded. However, this makes the numerical computation more demanding.

The modification (\ref{indcmatr}) is exact at the RPA level. Namely, the benzene 4-point polarizability $L$ obtained by solving equations (\ref{mateqforL}--\ref{BSEH}), including the modification (\ref{indcmatr}), represents an exact RPA 4-point polarizability screened by graphene. However, if we want to calculate the substrate renormalized molecular 4-point polarizability beyond RPA, the modification (\ref{indc}) is no longer sufficient.

 For example, when graphene is absent, the electron-hole interaction can be mediated directly by $V$ or indirectly via molecular polarization $VLV$, as shown by the BSE-Fock kernel in Fig.~\ref{Fig7}. Introducing graphene requires the replacing of $V$ by $\tilde W$.  This induces extra electron-hole interaction channels, such as interaction via graphene polarization $\Delta W$ and interaction via mixed molecular-graphene polarization $VL\Delta W$, $\Delta W LV$ and $\Delta W L \Delta W$.  However, the higher multiplicity processes, e.g. $VL\Delta WLV$ or $\Delta W L \Delta W L \Delta W$, etc.\ may be safely neglected.

 When the perturbed system is a small zero-dimensional object, such as a molecule, and the substrate is large higher dimensional object, such as a surface, the dominant substrate induced electron-hole interaction is via single substrate polarization $\Delta W$, and all higher order processes are negligible. Since we use the modification (\ref{indc}), we actually do include higher contributions such as $VL\Delta W$, $\Delta W LV$ and $\Delta W L \Delta W$. these multiple processes could influence the electron-hole interaction if the perturbed system is of the same dimensionality as the substrate. In that case, both systems can form coupled modes  and the propagator of the screened interaction $W$ can no longer be separated into propagators of individual screened interactions.\cite{JPCM,SSC}

The impact of the modification (\ref{indc}) to the $G_0W_0$ exchange correlation self-energy (\ref{sigmafin}--\ref{speicka}) is similar to its impact on the BSE-Fock term. After applying the modification (\ref{indcmatr}) to 
(\ref{Sigmamatr}), we neglect the above mentioned higher order processes. This does not influence the result significantly because we find that the dominant substrate induced modification of molecular self-energy comes from the single substrate polarization term $\Delta W$, as shown in Ref.~\onlinecite{StevenLouie2006}. After applying the modification (\ref{indcmatr}) to the expressions (\ref{sigmafin}--\ref{speicka}), the bare exchange self energy includes an additional exchange self energy term which contains single polarization $\Delta W$   
\begin{equation}
\Delta\Sigma_i^X(\omega)=-\sum^{N}_{j=1}\Delta W^{ij}_{ij}(\omega-\varepsilon_j).
\label{graingx}
\end{equation}
It is very important to note that the modification (\ref{indcmatr}) is unable to generate the substrate induced correlation self energy term which contains the single polarization $\Delta W$. Such an induced correlation term may be defined as  
\begin{eqnarray}
\Delta\Sigma^{C}_i(\omega)=\sum^{\infty}_{j=1}\int^{\infty}_0d\omega' \frac{\Delta S_{ij}(\omega')}{\omega-\varepsilon_j-\omega'+i\eta},
\label{indcorr}
\end{eqnarray}
where 
\begin{equation}
\Delta S_{ij}(\omega)=-\frac{1}{\pi}\Imag\left\{ \Delta W^{ij}_{ij}(\omega)\right\},
\label{indcorrS}
\end{equation}
may represent a significant correction to the self energy and therefore must be included by hand.

 In conclusion, the introduction of the substrate requires modification of the bare Coulomb interaction defined as ({\ref{indc}).  However, at the same time, the self energy (\ref{sigmafin}) should be corrected to include the induced correlation term of (\ref{indcorr}) and (\ref{indcorrS}). Therefore, the only task is to calculate the matrix elements of the induced Coulomb interaction (\ref{indmatrel}). 

The first step is to perform a Fourier transform in the $xy$-plane   
\begin{equation}
\Delta{W}({\bf r},{\bf r}',\omega)=\sum_{{\bf G}_{\parallel}}e^{i{\bf G}_{\parallel}\brho}\int\frac{d{\bf Q}}
{(2\pi)^2}e^{i{\bf Q}(\brho-\brho')}\Delta{W}_{\textbf{G}_{\parallel}}(\textbf{Q},\omega,z,z'),
\label{propex}
\end{equation}
where $\brho=(x,y)$, ${\bf Q}=(Q_x,Q_y)$ is a two-dimensional wave vector and $\textbf{G}_{\parallel}$ are graphene reciprocal vectors in the $xy$-plane. 

In Fig.~\ref{Fig11} we see that the equilibrium 
benzene-graphene separation is $z_0 \approx 6\ a_0$.  In  Ref.~\onlinecite{Duncan2} it is shown that the centroid of the induced density (density induced by the external point charge) is at $z_{im} \approx 2\ a_0$ from the graphene center, as shown in Fig.~\ref{Fig10}. This means that charge fluctuations in benzene feel like the ``external'' graphene field in the region $z,z'>z_{im}$. 
This is the region where the graphene induced density is zero. In this region the spatial part of the Fourier transform (\ref{propex}) has the simple form \cite{Duncan2,wake,gra-spectra}
\begin{equation}
\Delta{W}_{\textbf{G}_{\parallel}}(\textbf{Q},\omega,z,z')=
D(\textbf{Q}+\textbf{G}_{\parallel},{\bf Q},\omega)e^{-|{\bf Q}+\textbf{G}_{\parallel}|z-Qz'}. 
\label{ukr}
\end{equation}

From (\ref{ukr}) we see that the exponential factor cuts the higher $\textbf{G}_{\parallel}$ components. Since the average benzene graphene separation is $z_0 \approx 6\ a_0$, it is sufficient to keep only the $\textbf{G}_{\parallel}=0$ component. This has the consequence that $\Delta W$ becomes isotropic in $\brho$ and ${\bf Q}$ space, and the Fourier transform of the graphene field in the benzene region can be written simply as 
\begin{equation}
\Delta{W}_{\bf G_{\parallel}}(Q,\omega,z,z')=D(Q,\omega)e^{-Q(z+z')} \delta_{{\bf G}_\parallel 0}. 
\label{ukr0}
\end{equation}

The graphene electronic excitation propagator $D(Q,\omega)$ contains the intensities of all (collective and single particle) electronic excitations in graphene. The details of the calculation of the propagator $D(Q,\omega)$ can be found in Refs.~\onlinecite{wake}, \onlinecite{Duncan2}, and \onlinecite{gra-spectra}.  Here we use the same parameters employed in the calculation of $D(Q,\omega)$, except that the response function $\chi_0$ is calculated using a $201\times201\times1$ Monkhorst-Pack special \textbf{k}-point mesh, in order to have a finer \textbf{Q}-point mesh.

After inserting (\ref{ukr0}) into the expansion (\ref{propex}) and then together with the expansion (\ref{pairfunexp}) in the definition of the matrix elements (\ref{indmatrel}), we obtain
\begin{equation}
\Delta{W}^{kl}_{ij}(\omega)=\frac{1}{V^2_{uc}}\int\frac{d{\bf Q}}{(2\pi)^2}D(Q,\omega)e^{-2Qz_0}F^j_i({\bf Q})[F^l_k({\bf Q})]^*,
\label{indmelfin}
\end{equation}
where the form factors are defined as 
\begin{equation}
F^j_i({\bf Q})=\sum_{{\bf G}}C^j_i({\bf G})I({\bf Q},{\bf G}),
\label{finfor}
\end{equation}  
and 
\begin{equation}
I({\bf Q},{\bf G})= 8(-1)^{n_z}\frac{\sinh[Q\frac{L}{2}]\sin[(Q_x+G_x)\frac{L}{2}]\sin[(Q_y+G_y)\frac{L}{2}]}{(Q-iG_z)(Q_x+G_x)(Q_y+G_y)}.
\label{last}
\end{equation}

The summation in (\ref{finfor}) is over benzene superlattice reciprocal vectors ${\bf G}=(G_x,G_y,G_z)$ where $G_x = \frac{2\pi n_x}{L}$, $G_y = \frac{2\pi n_y}{L}$, $G_z = \frac{2\pi n_z}{L}$ and $n_x,n_y,n_z \in \mathbb{Z}$. The two-dimensional ${\bf Q}$ integration in (\ref{indmelfin}) is performed using a $61\times61$ rectangular mesh and the cutoff wave vector $Q_C=0.3\ a_0$.

In conclusion, the calculation of the quasiparticle spectra, optical absorption and energy loss spectra of the deposited benzene can be performed following the same recipe as for the isolated benzene (as described in Sec.~\ref{benzene-izolated}) except that the Coulomb matrix elements have to be renormalized as (\ref{indcmatr}) and calculated by using expressions (\ref{indmelfin}--\ref{last}). Also, the correlation self energy has to 
be corrected by the term (\ref{indcorr}).

\section{Results and discussion}
\label{Results}

In this section we use the formalism developed in Sec.~\ref{Apptomol} to calculate the quasiparticle properties, energies, and spectra of excitons in benzene deposited on various substrates. We also compare our results with  available experimental data.

\subsection{Quasiparticle properties of benzene on a substrate}

The quasiparticle energies for gaseous benzene are calculated directly by using $G_0W_0$ scheme (\ref{sigmafin}--\ref{reziduum}), additionally corrected by (\ref{graingx}) and (\ref{indcorr}) 
for the deposited benzene. As explained before, in order to get accurate energy shifts in gaseous benzene we use 45 unoccupied states, i.e.\ $60$ benzene states in total.

\begin{table*}
\caption{Comparison of the benzene ionization and affinity energy with experimental results.}\label{Eion}
\begin{ruledtabular}
\begin{tabular}{l|ccccccccc|c|c|c}
\ &\multicolumn{10}{c|}{occupied}&\multicolumn{2}{c}{unoccupied}
\\ 
\            &\ &\ &\ &\ &\ &\ &\ &\ &\ &HOMO&LUMO&     
\\ 
\hline
\            &$(2a_{1g})^2$&$(2e_{1u})^4$&$(2e_{2g})^4$&$(3a_{1g})^2$&$(2b_{1u})^2$&$(1b_{2u})^2$&$(3e_{1u})^4$&$(1a_{2u})^2$&$(3e_{2g})^4$&$(1e_{1g})^4$&$(2e_{2u})^4$&$(1b_{2g})^2$    
\\ 
DFT          &-24.25&-21.41&-17.75 &-15.79 &-13.99 &-13.94 &-13.15&-12.08&-11.18&-9.33&-4.18&-0.54       
\\ 
$G_0W_0$    &-26.51&-22.83&-19.78&-17.0&-16.38&-14.17&-14.16&-12.44&-11.61&-9.44&1.12&-2.85       
\\ 
Exp.\footnote{All ionization energies are taken from Refs.~\onlinecite{Spectro4} and \onlinecite{Spectro2} and the electron affinity energy is taken from Ref.~\onlinecite{exp_affin}.}
             &-25.9
             &-22.8
             &-19.2
             &-17.04 
             &-15.77
             &-14.47
             &-14.0
             &-12.3
             &-11.7
             &-9.45
             &1.12&  
\\
\end{tabular}
\end{ruledtabular}
\end{table*}

Table \ref{Eion} shows a comparison between benzene ionization and affinity energies obtained experimentally and by using the $G_0W_0$ scheme. The experimental ionization energies are taken from Refs.~\onlinecite{Spectro4} and \onlinecite{Spectro2}, 
while the electron affinity energy is taken from Ref.~\onlinecite{exp_affin}. 

To compare our quasiparticle energies with experimental values we must first determine the exact vacuum level. However, at the DFT level, we were unable to obtain an accurate vacuum level.  For this reason, we aligned the energy of the LUMO state $2e_{2u}$ with the experimental affinity energy of $1.12$~eV, and shifted all other levels accordingly. As can be seen in Table \ref{Eion}, the result of this procedure is that all quasiparticle energies, incorrect in the DFT calculations, are now in satisfactory agreement with the experimental ionization energies. We also see that the renormalization of the HOMO level is very weak while the LUMO state is shifted as much as $5$~eV upward.  In fact, its energy becomes positive, indicating that this state is unbound.        

The benzene excitons are mostly composed of transitions between the occupied state $1a_{2u}$, doubly degenerate occupied state $1e_{1g}$, doubly degenerate unoccupied state $1e_{2u}$ and unoccupied state $1b_{2g}$, which form the benzene $\pi-\pi^*$ complex. The molecular orbitals corresponding to these states are shown in Fig.~\ref{Fig13}. The energies of the doubly degenerate occupied states $1e_{1g}$ and doubly degenerate unoccupied states $2e_{2u}$ define the benzene HOMO--LUMO gap.

\begin{figure}
\centering
\includegraphics[width=\columnwidth]{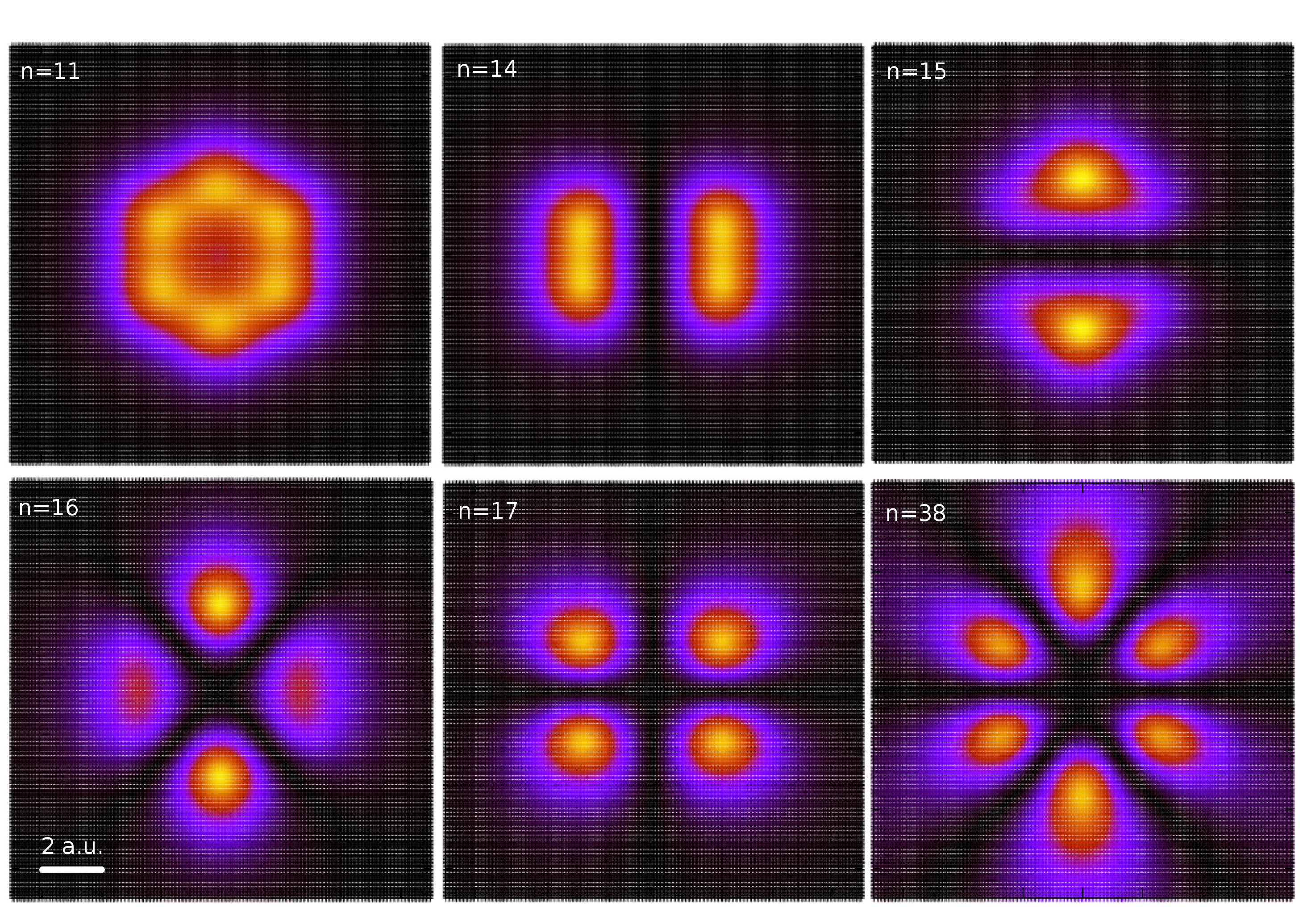}
\caption{(Color online) Molecular orbitals of the $\pi-\pi^*$ complex ($1a_{2u}$,$1e_{1g}$, $1e_{1u}$ and $1b_{2g}$) involved most dominantly in the formation of benzene excitons.}
\label{Fig13}
\end{figure}
 
\begin{table}
\caption{Quasiparticle HOMO--LUMO gaps for benzene deposited on various substrates.}
\label{G0W0}
\begin{ruledtabular}
\begin{tabular}{l|cccc}
 &Vacuum & \multicolumn{2}{c}{Graphene} & Ag\\
& &($\varepsilon_F=0$) & ($\varepsilon_F=1$~eV) & (jellium) 
\\
\hline
DFT &5.14 &5.05 & 5.14 & 5.14 
\\ 
$G_0W_0$&10.56 &8.55 &8.30 &8.22 
\\ 
Exp./Theor. &10.57\footnote{The experimental electron affinity energy is taken from Ref.~\onlinecite{exp_affin} and the ionization energy from Refs.~\onlinecite{Spectro4} and \onlinecite{Spectro2}.}
&7.35\footnote{Ref.~\onlinecite{StevenLouie2006}.}&\  &\  
\\
\end{tabular}
\end{ruledtabular}
\end{table}

Table \ref{G0W0} contains our $G_0W_0$ results for the HOMO--LUMO gap in benzene deposited on various substrates, as compared with available experimental and theoretical results. We can see that our approach for the HOMO--LUMO gap of benzene in gas phase is in excellent agreement with experimental results.  In this way we are able to verify the accuracy of our approach. 

Unfortunately, there are no available experimental results for benzene deposited on graphene or metal substrates. We find that all the substrates reduce the HOMO--LUMO gap approximately by the same amount, between $2$ and $2.4$~eV. This can be explained in terms of a simple 
image theory shift.  This can be obtained from (\ref{graingx}) and (\ref{indcorr}) by setting $\omega=\varepsilon_i$ and the summation index $j=i$. Then the induced self-energy becomes: 
\begin{equation}
\Delta\Sigma_i=\Delta\Sigma^X_i+\Delta\Sigma^C_i=
\left\{
\begin{array}{lr}
-\frac{1}{2}\Delta W^{ii}_{ii}(\omega=0);& i\leq N
\\
+\frac{1}{2}\Delta W^{ii}_{ii}(\omega=0);& i>N
\end{array}
\right..
\label{simim}
\end{equation} 

Since the static induced potential $\Delta W^{ii}_{ii}(\omega=0)$ is always negative, because positive/negative charges feel an attractive force from their negative/positive image charge. Thus, the degenerate  HOMO states $i=14,15$ are pushed up while the degenerate LUMO states $i=16,17$ are pushed down, reducing the HOMO--LUMO gap. The gap obtained using this simple model deviates less than $10\%$ from results obtained using the full expressions (\ref{graingx}) and (\ref{indcorr}).  This means that the dynamical effect only slightly corrects the simple image theory result (\ref{simim}). The same conclusion has been reached in Ref.~\onlinecite{StevenLouie2006} where the authors theoretically investigated the quasiparticle properties of benzene deposited on graphite. The benzene HOMO--LUMO gap is then reduced by 3.2~eV as shown in the last row of Table~\ref{G0W0}.

Considering the similar results obtained for pristine and doped graphene we expected the same result for graphite as well. Namely, the external charge is only able to induce charge in the surface region, i.e.\ charge in the first graphite monolayer, so for an external charge graphene should be the same as graphite.  Surprisingly, it instead turns out that the gap is about 1~eV larger than it is for graphite. Perhaps the graphite effective image plane $z_0$ is shifted outward compared to graphene.  This would strengthen the molecular orbital screening shift, and may explain the difference in the gap.

\subsection{Excitons in gaseous benzene}

\begin{table}
\caption{Comparison of the energy of the excitons in gaseous benzene with available experimental results.}
\label{benzenegas}
\begin{ruledtabular}
\begin{tabular}{l|ccccccc|cccccc}
\ &\multicolumn{7}{c|}{triplet}&\multicolumn{6}{c}{singlet}
\\ 
\hline
\            &\ &$B^3_{1u}$&\ &$E^3_{1u}$&\ &$E^3_{2g}$&\ &\ &$B^1_{2u}$&\ &$E^1_{1u}$&\ &$E^1_{2g}$     
\\ 
BSE(opt.abs.)&\ &3.93&\ &\ &\ &\ &\ &\ &\ &\ &7.02&\ &\       
\\ 
BSE(en.loss)&\ &3.93&\ &4.38&\ &6.81&\ &\ &4.80&\ &7.02&\ &8.55
\\ 
Exp.       &\ &3.95\footnote{Refs.~\onlinecite{Spectro3} and \onlinecite{triplet1_3_vacuum}.}
&\ &4.76\footnotemark[1]&\ &6.83\footnote{Refs.~\onlinecite{Spectro3} and \onlinecite{triplet4_vacuum}.} &\ &\ &4.90\footnote{Refs.~\onlinecite{Spectro3}, \onlinecite{triplet1_3_vacuum}, and \onlinecite{singlet_vacuum}.}&\ &6.94\footnotemark[3]&\ 
&7.80\footnotemark[3]
\\
\end{tabular}
\end{ruledtabular}
\end{table}

To ensure that our methodology for determining the energy of molecular excitons is accurate, we first calculate the energy of excitons in gaseous benzene.  This is easier to compare with the numerous experimental results which are available in the literature. 

The energies of the excitons in benzene in gas phase are determined from the positions of the peaks in the optical absorption spectrum calculated from expressions (\ref{apspowerF}) and (\ref{currf}).  Here the 4-point polarizability $L^{kl}_{ij}$ is obtained by solving the BSE (\ref{mateqforBSE}--\ref{BSEFOCK}). The incident electromagnetic wave is chosen to be $x$ polarized, i.e.\ in (\ref{incidentemp}) we set ${\bf e}=\hat{x}$. The energies of the excitons are also determined from the peaks in the energy loss spectrum calculated from expressions (\ref{energyloss2}), (\ref{Flossnew}) and (\ref{Ftcoef}). We chose an asymmetric external charge distribution so it can excite excitons of all symmetries.  The dipole is placed in the molecular plane but shifted by $4\ a_0$ in the $+x$ direction. The dipole is similarly polarized in $x$ direction, i.e.\ in (\ref{Ftcoef}) we put ${\bf p}=\hat{x}$ and ${\bf R}=4.0\hat{x}$. 

Table \ref{benzenegas} shows the energies of different excitons in gaseous benzene. To be consistent with available literature, we identified and labelled all excitons as shown in the first row of Table~\ref{benzenegas}.
Empty space in the table means that the corresponding spectra does not contain a corresponding exciton peak.  In other words, the exciton cannot be excited by a corresponding external driver. 

From  Table~\ref{benzenegas} we see that the triplet $B^3_{1u}$ and singlet $E^1_{1u}$ excitons are bright excitons, which can be excited by an electromagnetic field.  On the other hand, the triplet $E^3_{1u}$ and $E^3_{2g}$ and singlet $B^1_{2u}$ and $E^1_{2g}$ excitons are dark excitons, which cannot be excited by an electromagnetic field. This division to bright and dark excitons is consistent with optical and energy loss measurements.\cite{triplet1_3_vacuum,triplet4_vacuum,singlet_vacuum}  This means that our method simulates both classes of experiments well. In Table~\ref{benzenegas} we also see that the energies of all types of excitons, except for the dark $E^1_{2g}$ exciton, which is overestimated by $0.75$~eV, are in excellent agreement with the experimental data. Altogether, this suggests that the theoretical methodology we have developed works quite satisfactorily, and can be applied to molecules on substrates.

\subsection{Excitons in benzene on a substrate}

In this subsection we analyze the excitation spectra of benzene when the molecule is deposited on various substrates. In addition to the energy of excitons, special attention will be paid to the decay mechanism for excitons into real excitations within the substrate. In order to include these real excitations, we use full the dynamic BSE-Hartree kernel (\ref{BSEH}), (\ref{indcmatr}) and (\ref{indmatrel}), while the BSE-Fock kernel (\ref{BSEFOCK}) is also renormalized according to (\ref{indcmatr}) but remains static. Including the dynamical effects in the BSE-Hartree kernel causes the calculation of BSE to become very computationally demanding. However, we note that the transitions between three occupied ($n=11,14,15$) and three unoccupied ($n=16,17,18$) states forming the $\pi-\pi^*$ complex \cite{triplet1_3_vacuum,singlet_vacuum} participate most dominantly in forming all significant excitons in benzene. Therefore, we restrict our calculations to the transitions inside the $\pi-\pi^*$ complex.  This reduces the dimension of the BSE kernel matrix to only $18\times18$.

\begin{table}
\caption{Energy of excitons in benzene deposited on the various substrates. First row represents the optical absorption and second row the energy loss results. Data in parentheses represent the decay width $\Gamma$ of the corresponding exciton to electron-hole excitations in the substrate. Exciton energies are in ~eV and decay widths in meV.}
\label{benzex}
\begin{ruledtabular}
\begin{tabular}{l|ccc|ccc}
\ &\multicolumn{3}{c|}{triplet}&\multicolumn{3}{c}{singlet}
\\ 
\hline
\              &$B^3_{1u}$&$E^3_{1u}$&$E^3_{2g}$&$B^1_{2u}$&$E^1_{1u}$&$E^1_{2g}$ 
\\
Vacuum         &3.93&\ &\ &\ &7.02   
\\ 
\              &3.93&4.38&6.81&4.80&7.02&8.55   
\\\hline
Graphene   &4.08&\ &\ &\ &7.12\ (174)
\\ 
($\varepsilon_F=0$)            &4.08&4.37&7.15&4.80&7.12&8.89   
\\\hline
Graphene &4.10&\ &\ &\ &7.13\ (162) 
\\ 
($\varepsilon_F=1$~eV)           &4.10&4.38&7.18&4.81&7.13&8.93   
\\\hline
Ag       &4.10&\ &\ &\ &7.32\ (362)
\\ 
(jellium)              &4.11&4.38&7.19&4.81&7.31&8.93   
\\
\end{tabular}
\end{ruledtabular}
\end{table}

Table~\ref{benzex} shows energies of the different excitons after depositing on various substrates. For the separation between the molecular plane and the graphene plane or Ag jellium edge, we take the equilibrium value $z_0 \approx 6\ a_0$. 

It is interesting to note that the exciton energies are very weakly affected by the presence of the substrates. Similarly, the optical gap of benzene was previously found to be only weakly dependent on the height above a metal substrate.\cite{JuanmaRenormalization2} This seems to be a general property of the optical gap of weakly bound molecules on substrates.\cite{Catalin}

There are three dominant factors that define the exciton energy. First is the quasiparticle energy shift.  This changes the HOMO--LUMO gap and therefore the zero order exciton energy. Second is the fluctuation-fluctuation interaction present in BSE-Hartree kernel.  This increases the exciton energy. Third is the screened electron-hole interaction present in the BSE-Fock kernel.  This reduces the exciton energy. 

In our case, the substrate reduces the quasiparticle HOMO--LUMO gap by more than 2~eV, as shown in Table~\ref{G0W0}.  This reduces the exciton energy, and it barely influences the fluctuation-fluctuation interaction in the molecule. On the other hand, the substrate significantly weakens the electron-hole interaction, which increases the exciton energy. The latter can be explained by using simple image potential theory. When a substrate is present, the molecular electron interacts with the molecular hole but also with its negatively charged image.  This reduces the attractive electron-hole interaction and increases the exciton energy. Therefore, there is a competition between the HOMO--LUMO gap reduction and weakening of the electron-hole interaction. 

In our case, substrate induced electron(LUMO)-hole(HOMO) interaction, which can be approximated by the matrix elements $-\Delta W^{jj}_{ii}(\omega=0)>0$, almost exactly cancels the HOMO--LUMO gap reduction.  Using (\ref{simim}), this can be approximated as 
\begin{equation}
\Delta\Sigma_{\textrm{LUMO}}-\Delta\Sigma_{\textrm{HOMO}}=
\frac{1}{2}\left[\Delta W^{jj}_{jj}(\omega=0)+\Delta W^{ii}_{ii}(\omega=0)\right]<0, 
\end{equation}
where  $i=14,15$ and $j=16,17$. Therefore, the exciton energies are indeed substantially affected by different mechanisms.  However, these mechanisms cancel each other and the exciton energy remains almost unchanged.

The theoretical model developed here allows us to analyze the molecule/substrate spectra in analogy to the spectra of a driven/damped harmonic oscillator. Namely, the calculation is performed in such a way that the external electromagnetic wave or dipole (driving force) can induce current or charge in the molecule (harmonic oscillator) but not in the substrate (damping source). However, the molecule interacts with the substrate and it can excite plasmons (leading to extra peaks in the molecular spectra) or electron-hole excitations (influencing final exciton width) in the substrate. The inverse exciton width represents the decay rate of the initially excited exciton. 

Here we note that the molecule is a zero-dimensional object.  This means there is no translational invariance within it, and ${\bf Q}$ is not a valid quantum number. This also means that the exciton at fixed frequency $\omega$ can decay into any of the electron-hole excitations with any momentum transfer ${\bf Q}$. Since we use the static BSE-Fock kernel in our theoretical model, triplet exciton are always sharp peaks and cannot decay into substrate excitations.  For this reason, we investigate the decay of singlet excitons. 

To be able to distinguish a substrate induced exciton decay from intrinsic decay, we choose an intrinsic exciton decay which is very small, namely $1$~meV. 
Since the interaction between the molecule and substrate is quite weak, the probability of transitions between the $E_{1u}$ exciton of benzene and the substrate is well described by Fermi's Golden Rule \cite{FermisGoldenRule}.  As the substrates considered are in the wide-band limit, the final density of states is a Lorentzian distribution, so that the absorption spectra has the form
\begin{equation}
A(\omega) \approx \frac{2|H|^2}{\hslash}\frac{\Gamma/2}{(\omega-\omega_0)^2+(\Gamma/2)^2}.
\label{lor}
\end{equation}
Here $H$ is the coupling matrix element between the molecule and substrate, $\omega_0$ is the exciton energy,  and $\tau \approx \hslash/\Gamma$ is its lifetime.\cite{DuncanJPCLett}  Assuming $|H|^2$ is only weakly dependent on the energy, we may fit (\ref{lor}) to the calculated spectra to estimate the inverse lifetime of the exciton, i.e.\ the decay width $\Gamma$.

\begin{figure}
\centering
\includegraphics[width=1.\columnwidth]{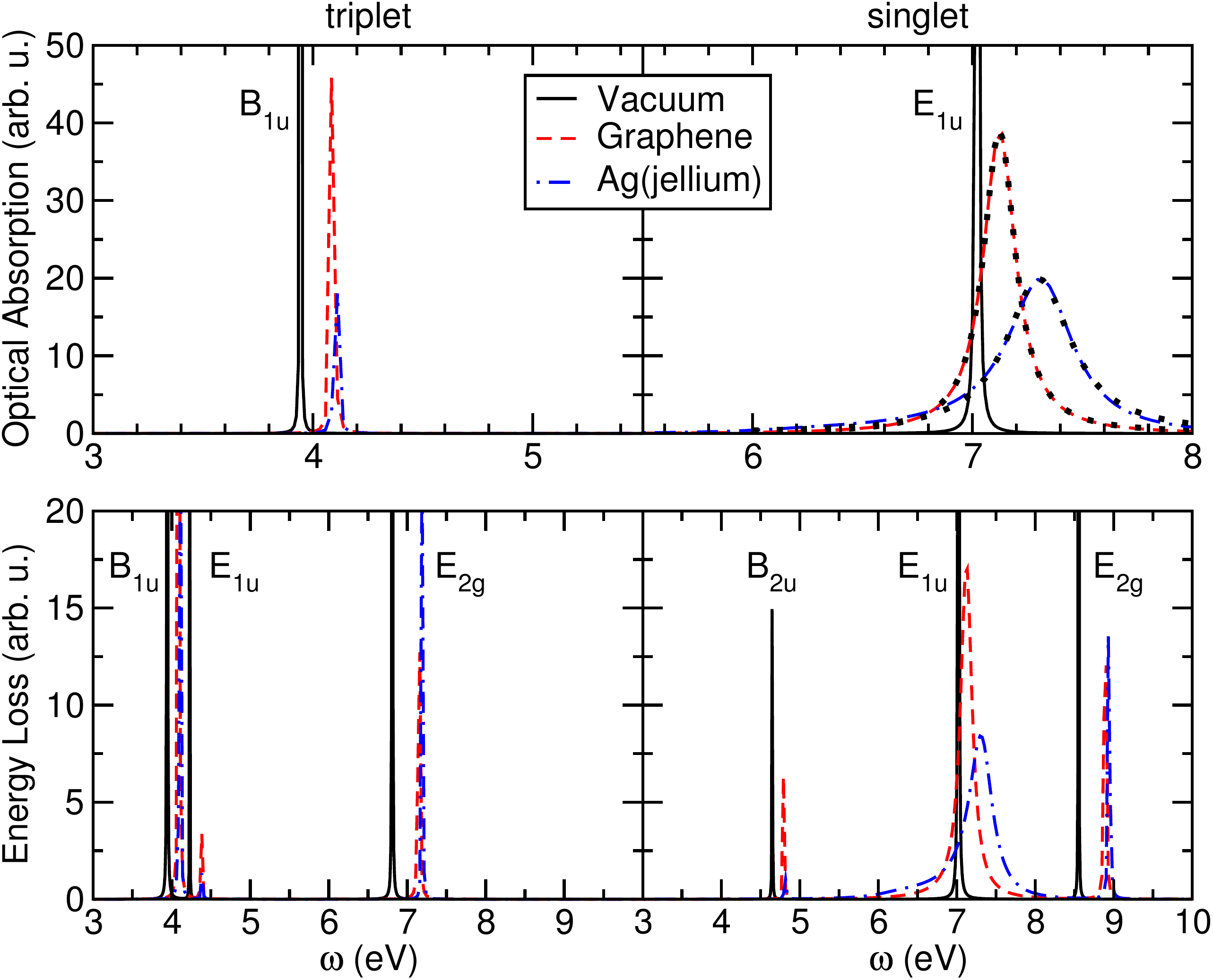}
\caption{(Color online) Optical absorption and energy loss spectra of benzene in vacuum (black solid line), on pristine graphene (red dashed line), and in the vicinity of a Ag(jellium) surface (blue dash-dotted line). For separation between molecular plane and graphene plane or jellium edge we take the equilibrium distance $z_0 \approx 6.0\ a_0$. Black dotted lines are Lorentzian fits to the $E_{1u}$ optical adsorption spectra.}
\label{Fig14}
\end{figure}

Fig.~\ref{Fig14} presents the optical and energy loss spectra of gaseous benzene (black solid line) and of  benzene deposited on pristine graphene (red dashed line). Upper panels show the optical absorption spectra showing only those excitons that can be excited by the external electromagnetic wave. This is why we can notice the absence of triplet $E^3_{1u}$ and $E^3_{2g}$ excitons and singlet $B^1_{2u}$ and $E^1_{2g}$ excitons. Lower figures show the spectra of excitons excited by a dipole. These spectra show all types of dark and bright excitons. We see that the substrates do not generate any extra peak in the spectra.  This means that the excitons do not interact with the $\pi-\pi$ plasmons in pristine graphene. Also, as mentioned previously, we see that the exciton energies are almost unaffected by the substrate and that they are just slightly shifted toward higher energies. From energy loss spectra (lower panels of Fig.~\ref{Fig14}) we see that only the singlet $E^1_{1u}$ excitons obtain a final width, while the dark $B^1_{2u}$ and $E^1_{2g}$ excitons remain sharp.

By observing the low momentum/energy ($Q\le 0.3\ a_0$ and $\omega<10$~eV) graphene spectra \cite{gra-spectra} we can see a wide interband $\pi-\pi^*$ electron-hole continuum and broad $\pi-\pi^*$ plasmon. So obviously the $E^1_{1u}$ exciton decays into all types of $\pi-\pi^*$ excitations in graphene.  This does not exclude the possibility that it can also interact with the weak graphene $\pi-\pi^*$ plasmon. By fitting $E^1_{1u}$ optical spectra to a Lorentzian (\ref{lor}), we obtain $\Gamma \approx 174$~meV as provided in Table~\ref{benzex}.

The spectra of benzene deposited on doped graphene ($\varepsilon_F=1$~eV) is almost identical to that for pristine graphene and is not shown here. This is expected because even though the doping induces extra intraband electron-hole continuum, it appears at energies ($\omega<2$~eV) \cite{gra-spectra} lower than the benzene exciton energies ($\omega_0>3$~eV). There is also an absence of extra peaks in the spectra.  This means that the excitons do not interact with the 2D plasmon of doped graphene. The decay width of the $E^1_{1u}$ exciton is slightly reduced in the vicinity of doped graphene to $\Gamma \approx 162$~meV. This could be because in the doped graphene the interband $\pi-\pi^*$ continuum is shifted to higher energies.  This reduces their intensity at the exciton energy $\omega_{E^1_{1u}}\approx 7$~eV. From this we can conclude that the excitons in larger molecules, as e.g.\ in terrylene $\text{C}_{30}\text{H}_{16}$ where $\omega_{0}\leq 3$~eV, would be more strongly influenced by graphene doping.

Fig.~\ref{Fig14} also shows the optical and the energy loss spectra of benzene deposited on a Ag (jellium) surface (blue dashed-dotted line). For the separation between molecular plane and jellium edge we chose $z_0 \approx 6\ a_0$. We see that the singlet $E^1_{1u}$ exciton significantly decays into excitations within the metal. By fitting its optical spectrum to a Lorentzian, we obtain for its width $\Gamma \approx 362$~meV. This is because in the jellium metal there are many interband electron-hole channels into which it can decay. It is interesting to note that even in this case, when the phase space of the electron-hole excitation becomes very rich, the dark excitons ($B_{2u}$ and $E_{2g}$ in the lower panels of Fig.~\ref{Fig14}) remain sharp. 

We conclude that only the bright exciton $E^1_{1u}$ decays into real electron-hole excitations in the substrate.  On the other hand, the dark excitons $B^1_{2u}$ and $E^1_{2g}$ do not interact with any type of real excitations in semi-metallic or metallic substrates and remains in well defined eigenmodes.

\section{Conclusions}
\label{Conclusions}
In this manuscript we presented how the molecular optical and energy loss spectra can be obtained directly from the dynamical 4-point polarizability 
matrix $L^{kl}_{ij}(\omega)$, which is the solution of the BS matrix equation. We demonstrated that the inclusion of substrates requires minimal intervention to the 
presented formulation.  This implies that everywhere throughout the BSE-$G_0W_0$ scheme the bare Coulomb interaction $V$ can be replaced by the dynamically screened Coulomb interaction $W(\omega)=V+\Delta W(\omega)$, where $\Delta W(\omega)$ is the substrate induced Coulomb interaction. 

This formulation has been successfully applied to the calculation of the quasiparticle energy levels and exciton energies in the isolated benzene molecule. The method has then been applied to calculate the electronic structure and excitations in benzene deposited on pristine and doped graphene 
and in benzene in the vicinity of a Ag(jellium) surface. It is shown that the substrates cause a reduction of the quasiparticle HOMO--LUMO gap (by about $2$~eV), which weakly depends on the type of substrate. We have also shown that the energy of all excitons in the isolated molecule remain relatively unchanged when the molecule is deposited on a substrate. 

By using an image theory based argument, we note that the exciton energies are under the influence of two mechanisms which tend to cancel each other out.
The substrate reduces the quasiparticle HOMO--LUMO gap which reduce exciton energy.  However, at the same time, the induced image-electron or image-hole weakens the electron-hole interaction which raises the exciton energy. 

We pay special attention to the investigation of the interaction of different types of excitons with real electronic excitations in the substrate. It is noted that only the optically active $E_{1u}$ exciton decays into the electron-hole excitations in the 
substrates.  However, it does not couple to any plasmons in doped graphene or within the metallic surface. 

Coupling to electronic excitation in the substrate 
causes a Lorentzian broadening of the $E_{1u}$ exciton, whose  width is $\Gamma \approx 174$~meV for pristine graphene and $\Gamma \approx 362$~meV for metal 
surfaces as substrate. We have also noticed that the exciton quenching could be tuned by graphene doping. 

Although this effect is not observed 
in benzene, it should exist for larger $\pi$-conjugated complexes, such as terrylene $\text{C}_{30}\text{H}_{16}$.  There the molecular 
excitons falls in the gap between upper and lower edges of the doped graphene intra and interband electron hole continuum respectively.
Now that the developed formulation has been successfully tested, it has the potential to be applied to more computationally demanding and technologically interesting molecular systems.   

\acknowledgments
V.D.\ is grateful to the Donostia International Physics Center (DIPC) and Pedro M. Echenique for their hospitality during various stages of this 
research. D.J.M.\ acknowledges funding through the Spanish ``Juan de la Cierva'' program (JCI-2010-08156), Spanish Grants (FIS2010-21282-C02-01) and (PIB2010US-00652), and ``Grupos Consolidados UPV/EHU del Gobierno Vasco'' (IT-578-13). The authors also thank M.~\v Sunji\' c and I. Kup\v ci\' c for useful discussions.

\end{document}